\documentclass[preprint,3p,times,sort&compress]{elsarticle}

\usepackage[activate={true,nocompatibility},final,tracking=true,kerning=true,spacing=true,factor=1100,stretch=10,shrink=10]{microtype}
\microtypecontext{spacing=nonfrench}
\usepackage[T1]{fontenc}
\usepackage{amssymb}
\usepackage{amsmath}
\usepackage{bm}
\usepackage{hyperref}
\usepackage{siunitx}
\usepackage[capitalize]{cleveref}
\usepackage[english]{babel}
\usepackage[section]{placeins}
\usepackage{booktabs}
\usepackage{xcolor}
\usepackage{verbatim}

\usepackage{tikz}

\DeclareRobustCommand{\linesym}[1]{%
  \tikz[baseline=-0.6ex]\draw[#1,line width=0.6pt] (0,0)--(1.2em,0);%
}

\renewcommand{\vec}[1]{\bm{#1}}
\newcommand{\up}[1]{\mathrm{#1}}
\renewcommand{\bar}[1]{\mkern3.0mu\overline{\mkern-3.0mu#1}\mkern1mu}

\journal{Nuclear Physics B}

\begin{document}
\begin{samepage}

    \begin{frontmatter}
        \title{Non-equilibrium scaling across first-order transitions with relativistic scalar fields}

        \author[JLU]{Leon J. Sieke\corref{cor1}}
        \ead{leon.j.sieke@physik.uni-giessen.de}
        \cortext[cor1]{Corresponding author}

        \author[JLU]{Jessica Fuchs}
        \ead{jessica.fuchs@physik.uni-giessen.de}

        \author[JLU,HFHF]{Lorenz von Smekal}
        \ead{lorenz.smekal@physik.uni-giessen.de}

        \affiliation[JLU]{organization={Institut f\"ur Theoretische Physik, Justus-Liebig-Universit\"at},
            addressline={Heinrich-Buff-Ring 16},
            city={35392 Gießen},
            country={Germany}}

        \affiliation[HFHF]{organization={Helmholtz Forschungsakademie Hessen f\"ur FAIR (HFHF)},
            addressline={Campus Gießen},
            city={35392 Gießen},
            country={Germany}}

        \begin{abstract}
            We investigate the out-of-equilibrium dynamics of a relativistic $Z_2$-symmetric scalar field theory with Langevin dynamics in two and three spatial dimensions under linear driving across magnetic first-order phase transitions, close to and far below the critical temperature $T_c$.
            Using classical-statistical lattice simulations, we find that if the driving timescale is sufficiently fast, the system exhibits finite-time scaling behavior independent of temperature and dimensionality, identical to that observed in mean-field simulations.
            In slow quenches near $T_c$ this mean-field behavior crosses over to critical Kibble-Zurek scaling behavior, while for temperatures $T \ll T_c$ nucleation and growth dominate the transition dynamics, resulting in corrections to scaling.
            Near the transition point where the order parameter changes sign, the crossover between mean-field and critical out-of-equilibrium dynamics is found to be well described by the leading algebraic correction to Kibble-Zurek scaling.
            We find that universal non-equilibrium scaling behavior can be observed for $T \lesssim T_c$, provided the driving is fast enough to avoid nucleation but slow enough for correlations to form, and compute the associated universal scaling functions for the order parameter.

        \end{abstract}


    \end{frontmatter}

    \tableofcontents
\end{samepage}

\vfill

\section{Introduction}
First-order and critical phase transitions behave qualitatively different under nearly adiabatic conditions.
The former proceed through nucleation and growth, and are accompanied by metastability and hysteresis, while the latter feature diverging correlations and universal scaling.
When transitions occur in finite time, non-equilibrium effects gain relevance, and this sharp distinction becomes blurred.
The dynamic properties of systems undergoing continuous phase transitions near a critical point have been extensively studied in the literature. The behavior in temperature-driven cooling transitions is well described by the established Kibble-Zurek mechanism~\cite{Kibble:1976sj,Zurek:1985qw,Zurek:1996sj}, which relates the density of defects formed during the transition to the quench rate via universal scaling laws.
Similarly, a generalized theory of finite-time scaling~\cite{Gong:2010,Zhong:2011,Huang:2014oma,Huang:2016,Feng:2016qmw} is able to capture the dynamic hysteresis and scaling behavior observed in field-driven phase transitions across the critical point in terms of critical exponents and universal scaling functions.
These theories  have since been successfully applied to a variety of classical and quantum systems~\cite{Bowick:1992rz,Laguna:1997,Laguna:1998,Yates:1998kx,Antunes:1998rz,Monaco:2002zz,Zurek:2005kod,Dziarmaga:2005efq,Polkovnikov:2005xfc,Polkovnikov:2007xy,Damski:2007jy,Dziarmaga:2009,DeGrandi:2011vmo,Kolodrubetz:2012vwl,Chandran:2012cjk,delCampo:2013nla,Liu:2013nla,Lamporesi:2013,Navon:2014bjr,Yin:2014,Mukherjee:2016kyu,Weiss:2018uoa,Keesling:2018ish,Puebla:2019rbi,Weinberg:2020mba,Yuan:2021_feb,Yuan:2021fbk,Yuan:2021_jul,King:2022okf}.

Scaling behavior and out-of-equilibrium dynamics at first-order phase transitions have also been studied extensively in the literature for classical and quantum systems alike~\cite{Nienhuis:1975zs,Fisher:1982xt,Binder:1987,Rao:1990,Jung:1990,Lo:1990,Sengupta:1992,Somoza:1993,Thomas:1993,Luse:1994,Rikvold:1994za,Zhong:1994,Zhong:1995,Zhong:1995_PRL,Sides:1998,Sides:1999,Chakrabarti:1999,Zhong:2002_PRB,Zhong:2005_PRL,Loscar:2009,Amin:2009bef,Jorg:2010bne,Fan:2011,Zhong:2012tb,Berganza:2014,Panagopoulos:2015uia,Pelissetto:2016,Pelissetto:2016tvy,Liang:2016zok,Pelissetto:2017lxo,Panagopoulos:2018hva,Zhong:2018,Bar:2018,Scopa:2018,Pelissetto:2018mmz,Pelissetto:2018,Iino:2018rox,Fontana:2019zgc,Qiu:2020,Pelissetto:2020,Rossini:2021swu,Sinha:2021aua,Tarantelli:2023kos,Pelissetto:2024,Zhong:2024baf,Zhang:2025_JPSJ,Pelissetto:2025fdb,Pelissetto:2025,Zhong:2025_CPL,Zhang:2025_CPL,Pelissetto:2025pzi,Pelissetto:2026gzg}.
Although many of these works reported power-law scaling in the hysteresis behavior with respect to the quench rate, the question of whether this scaling is universal is more controversial, as many of the reported scaling exponents have since been shown to be parameter dependent and only effective~\cite{Mahato:1993,Zhong:2024baf}.
Furthermore, the mechanism governing the first-order transition dynamics differs qualitatively between systems of finite size, and in the thermodynamic limit, adding additional subtleties to the interpretation of scaling behavior~\cite{Rikvold:1994za,Pelissetto:2025pzi}.
This has resulted in the Kibble-Zurek dynamics for first-order transitions being generally less well understood than for their critical counterparts.

It has recently been demonstrated that universal exponents emerge once additional non-universal contributions to an effective cubic theory are properly taken into account~\cite{Zhong:2024baf,Zhang:2025_JPSJ,Zhong:2025_CPL,Zhang:2025_CPL}.
However, robust scaling collapse according to this theory is only achieved when additional parameters such as the strength of interactions, which are usually irrelevant for critical behavior, are properly scaled with the quench rate, otherwise the exponents found via usual scaling methods will only be effective~\cite{Zhong:2024baf}.
In systems where interactions are not easily adjustable, as is the case for most real systems in experiments, probing this kind of universal scaling behavior may then be much more difficult than for conventional critical dynamics.

In this work, we rather focus on non-equilibrium phase transitions at fixed interaction strengths.
We build upon a previous study of field-driven phase transitions across the critical point~\cite{Sieke:2024dns}, and extend it to the case of linear quenches across the first-order line.
Close to the critical point, the dynamics of the system are still expected to be governed by critical scaling-laws and universal  two-variable scaling functions describing the deviations from pure Kibble-Zurek scaling due to the finite distance from the critical point.
According to the finite-time scaling theory~\cite{Zhong:2011}, these deviations become small in sufficiently fast quenches.
However, in very rapid quenches, Kibble-Zurek scaling is known to break down in the case of temperature-driven transitions~\cite{Zeng:2022hut,Rao:2025xku}.
Here, we will investigate under which conditions critical non-equilibrium scaling can arise in field-driven first-order transitions, how this scaling breaks down in rapid quenches, and when the traditional dynamics of nucleation and growth  dominate the transition.

We find, in accordance with the finite-time scaling theory~\cite{Zhong:2011}, that universal non-equilibrium scaling behavior can be observed if the dimensionless scaling variable $\tau / R^{1/\nu r_c}$ is small, or equivalently if the quench rate $R$ is sufficiently large, i.e. $R \gg \tau^{\nu r_c}$ for a given reduced temperature $\tau = (T-T_c)/T_c$, where $r_c = z + \beta\delta/\nu$ is a combination of critical exponents.
In rapid quenches however, we find that the behavior of the system becomes independent of temperature and dimensionality, and approaches a mean-field description.
In this regime, finite-time scaling of the order parameter with the  quench rate can still be observed, when using mean-field exponents instead of the critical ones.

This work is structured as follows. In \cref{sec:model_setup}, we introduce our model, namely a relativistic $Z_2$-symmetric scalar field theory, and the quench protocol used to drive the system through the first-order phase transition.
In \cref{sec:theories}, we briefly summarize relevant theories for the driven first-order phase transitions we are investigating, first and foremost the theory of finite-time scaling which describes the universal scaling behavior near the critical point, theories of nucleation and growth that are expected to govern the dynamics in slow quenches at low temperatures, followed by a discussion of the mean-field theory which we find to accurately describe the dynamics in rapid quenches.
In \cref{sec:mean_field_results}, we first present our numerical results obtained from mean-field simulations of the model, setting the stage for the subsequent discussion of the results for the fully fluctuating theory obtained from classical-statistical lattice simulations in $d=2$ and $d=3$ spatial dimensions that follows in \cref{sec:numerical_results}.
Finally, we conclude in \cref{sec:conclusions} with a summary of our findings and highlight possible future research directions.

\section{Model setup}
\label{sec:model_setup}
We study a system in the static $Z_2$ Ising universality class and the dynamic universality class of Model A in the Hohenberg-Halperin classification scheme~\cite{Hohenberg:1977ym}, characterized by a nonconserved real scalar order-parameter field $\phi $ following the second-order Langevin equation of motion
\begin{equation}
    \frac{\partial^2\phi}{\partial t^2} + \gamma \frac{\partial \phi}{\partial t} = -\frac{\delta F[\phi]}{\delta \phi} + \xi.
    \label{eq:eom}
\end{equation}
Here $\xi$ is a stochastic noise term with zero mean $\langle \xi \rangle = 0$ and variance $\langle \xi(t, \vec{x})\xi(t', \vec{x'})\rangle = 2\gamma T\delta({t-t'})\delta({\vec{x}-\vec{x'}})$, where the temperature $T$ determines the strength of fluctuations, and $\gamma$ is a damping coefficient.
In the case of strong damping or long timescales, where $\gamma \left|\frac{\partial \phi}{\partial t}\right| \gg \left|\frac{\partial^2\phi}{\partial t^2}\right|$, the inertial term may be neglected, and \cref{eq:eom} reduces to the overdamped Langevin equation used in~\cite{Hohenberg:1977ym} to define Model A dynamics. Both variants of the equation of motion capture exactly the same universal critical properties, since they are determined by the slow long-wavelength modes of the system, which are not affected by the inertial term.

The Landau-Ginzburg-Wilson (LGW) free-energy functional $F$ in our model in $d=2,3$ spatial dimensions is given by
\begin{equation}
    F[\phi] = \int d^dx \left\{ \frac{1}{2}\left(\nabla \phi \right)^2 + f(\phi) \right\},\quad \text{with } f(\phi) = \frac{m^2}{2} \phi^2 + \frac{\lambda}{4!}\phi^4 - J \phi.
    \label{eq:free_energy}
\end{equation}
The equilibrium partition function then reads
\begin{equation}
    Z_\text{eq} = \int \mathcal{D}\phi \, \up{e}^{-F[\phi]/T}.
\end{equation}

In the absence of an external symmetry breaking field, i.e. $J=0$, the model is symmetric under the $Z_2$ transformation $\phi \to - \phi$. For $m^2 <0 $, this symmetry is spontaneously broken below a critical temperature $T<T_c$, where the phase transition changes from continuous to first-order and two ordered phases with opposite sign of the order parameter coexist.

To solve \cref{eq:eom} numerically with classical-statistical lattice methods \cite{Aarts:2001yx,Berges:2009jz,Schlichting:2019tbr,Schweitzer:2020noq,Schweitzer:2021iqk}, we discretize the system on a square (cubic) lattice of volume $V=L^d$ with periodic boundary conditions in $d=2$ and $3$ spatial dimensions.
The numerical setup closely follows that of Refs.~\cite{Schweitzer:2020noq,Sieke:2024dns}.
The lattice spacing is set to unity throughout this work and all quantities are expressed in corresponding lattice units. In particular, as in Refs.~\cite{Schweitzer:2020noq,Sieke:2024dns}, the model parameters are set to $m^2 = -1$, $\lambda = 1$ and $\gamma=0.1$ in these units throughout this work.

The order parameter, which will be referred to as the magnetization in the following, is defined by the spatial average of the order-parameter field (at time $t$),
\begin{equation}
    M(t) = \frac{1}{V} \sum_{\bm{x}} \phi_{\bm{x}}(t).
\end{equation}
Any observable is then obtained as an average over the classical-statistical ensemble of many independent simulation runs with different random realizations of the stochastic noise $\xi$.

\begin{figure}[tb]
    \centering
    \includegraphics[width=.357\linewidth]{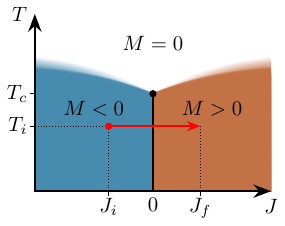}
    \caption{\label{fig:phase_diagram} Qualitative phase diagram of the model in the $J$-$T$ plane. The system is prepared in a negatively magnetized state
        at $T_i<T_c,\ J_i < 0$ and quenched to $J_f = -J_i$ at constant temperature with a constant rate $R$ across the first-order phase transition line (quench trajectory represented by the red arrow).
        The critical point located at $T=T_c,\ J=0$ is indicated by the black dot.
        The ordered phases with negative and positive magnetization are shown in blue and orange shading respectively, while the disordered phase is represented by white shading.
        The first-order transition line at which the two ordered phases can coexist is shown as a solid black line.
    }
\end{figure}

We investigate the dynamics of the system when driven through the first-order phase transition by varying the external field $J$ at a constant temperature $T<T_c$. To this end, a linear quench protocol is employed, where the external field $J$ is varied
from an initial value $J(t_i) < 0$ to a final value $J(t_f) = -J(t_i) > 0$ at a constant rate $R>0$ according to
\begin{equation}
    J(t) = R t,
\end{equation}
where the time $t$ is normalized so that the point $J=0$ is always reached at $t=0$. A schematic illustration of the phase diagram and quench trajectory is shown in \cref{fig:phase_diagram}.
The initial state is prepared at $J_i \equiv J(t_i)< 0$ by equilibrating the system at constant temperature $T < T_c$ using direct numerical simulation of the Langevin \cref{eq:eom} with $J=J_i$ for a sufficiently long thermalization period. In principle, more efficient Markov chain Monte-Carlo algorithms can be used to speed up the generation of initial conditions. This was not necessary here, as the initial state is prepared away from the critical point and the thermalization period is negligibly short compared to the driving timescales considered in this work.

\section{Theories for driven first-order phase transitions}
\label{sec:theories}

The way in which the system transitions from one ordered phase to the other depends strongly on the temperature and the rate at which the external field is changed.
In the extreme case of very slow quenches far below the critical temperature ($T\ll T_c$), the system evolves quasi-adiabatically and follows the equilibrium magnetization curve closely until the coexistence region at $J=0$ is reached, where the initial phase becomes metastable.
The transition is then initiated locally via nucleation of bubbles of the new phase, which subsequently grow and coalesce until the system has fully transitioned to the new ordered phase.
The finite value of the external field at which the overall magnetization changes sign, is referred to as the \emph{coercive field} $J_c$ and generally depends on the temperature and rate of quenching.

\begin{figure}[tb]
    \centering
    \begin{minipage}{.48\textwidth}
        \centering
        \scriptsize{$\tau = 0$}\\[.1cm]
        \includegraphics[width=\textwidth]{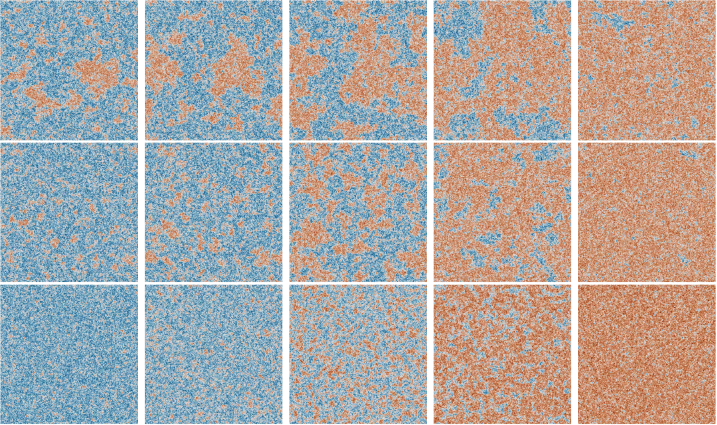}
    \end{minipage}
    \hfill
    \begin{minipage}{.48\textwidth}
        \centering
        \scriptsize{$\tau = -0.2$}\\[.1cm]
        \includegraphics[width=\textwidth]{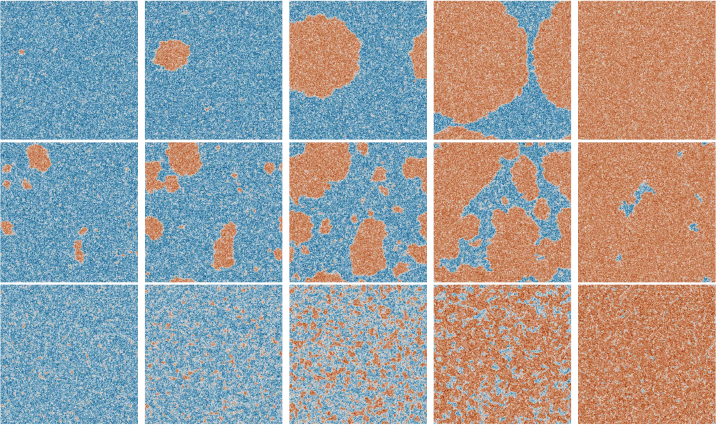}
    \end{minipage}

    \caption{\label{fig:field_configurations}
        Snapshots of field configurations during the phase transition across the critical point ($\tau = 0$, left) and the first-order transition line ($\tau = -0.2$, right) for quench rates $R \in \{10^{-6}, 10^{-4}, 10^{-2}\}$ (top to bottom). Negative (positive) values of the order parameter field $\phi$ are shown in blue (orange). Snapshots are taken at equally spaced values of the external field $J$: for $\tau = 0$, at $J \in \{5, 10, 15, 20, 25\} \times 10^{-4}$, $\{5, 10, 15, 20, 25\} \times 10^{-3}$, and $\{5, 10, 15, 20, 25\} \times 10^{-2}$ (top to bottom); for $\tau = -0.2$, at $J \in \{2.45, 2.50, 2.55, 2.60, 2.65\} \times 10^{-2}$, $\{5, 10, 15, 20, 25\} \times 10^{-2}$, and $\{15, 20, 25, 30, 35\} \times 10^{-2}$ (top to bottom).}
\end{figure}

In the case of faster quenches that come sufficiently  close to the critical point (for $T\lesssim T_c$), the system cannot follow the changing external field arbitrarily due to critical slowing down and consequently falls out of equilibrium before reaching the coexistence line.
In this case, the system can be driven through the region of metastability faster than nucleation can occur, until the initial phase eventually becomes unstable.
The system can then transition from one ordered phase to the other one on a global scale without the need for fluctuations initiating nucleation.
Snapshots of typical field configurations in quenches across the critical point and the first-order transition line for different quench rates are shown in \cref{fig:field_configurations}.

Hysteresis phenomena are observed in all cases, where the magnetization lags behind the changing external field. However, the details of when and how the phase transition proceeds, differ significantly between the different regimes.
We therefore separate the following discussion into three parts, first addressing near-critical quenches where observables can be described in terms of universal critical exponents and scaling functions.
Then follows a brief review of classical nucleation and growth, relevant for slow quenches far below the critical temperature.
Finally we provide a short overview of the mean-field approximation which, as we will see, describes the  behavior of systems  in rapid quenches very well, where strong fields are reached rendering fluctuations irrelevant for the transition mechanism.

\subsection{Near critical phase transitions: Kibble-Zurek mechanism and finite-time scaling}
In trans-critical quench protocols, where the system is driven exactly across the critical point at a finite rate $R$, the system is guaranteed to fall out of equilibrium before reaching the critical point, due to critical slowing down.
The dynamic behavior of observables, e.g. the magnetization, is then to leading order described by the theory of finite-time scaling~\cite{Gong:2010,Huang:2014oma,Feng:2016qmw}, a generalization of the well established Kibble-Zurek mechanism~\cite{Kibble:1976sj,Zurek:1985qw,Zurek:1996sj}.
Accordingly, due to scale invariance near the critical point, the magnetization $M$, as any other observable in general, follows a scaling relation of the following form~\cite{Zhong:2011}
\begin{equation}
    M(\tau, J, R) = b^{-\beta/\nu} M\left((-\tau) b^{1/\nu}, J b^{\beta\delta/\nu}, R b^{r_c}\right), \quad \tau < 0,\ J>0,\ R > 0
    \label{eq:scaling_ansatz}
\end{equation}
where $b$ is an arbitrary length-rescaling factor, $\tau = (T-T_c)/T_c$ is the reduced temperature, and $r_c = z + \beta\delta/\nu$ is a combination of static critical exponents $\nu$, $\beta$, $\delta$, and the dynamic critical exponent $z$.
To obtain \emph{universal} scaling functions, we normalize all dimensionful quantities by their non-universal amplitudes and express all scaling functions in terms of the corresponding dimensionless variables
\begin{equation}
    \bar{J} \equiv J / J_0 = J (B^c/B)^\delta, \quad \bar{M} \equiv M / B, \quad \bar{R} \equiv R / R_0, \quad \tau \equiv (T-T_c)/T_c.
\end{equation}
The critical amplitudes of our model have been determined in previous works~\cite{Schweitzer:2020noq,Sieke:2024dns} and are summarized in \cref{tab:critical_amplitudes}.

As our model setup is not optimized to determine critical exponents with high precision, we will use established literature values for the static critical exponents of the 2D and 3D Ising universality classes, as well as the dynamic critical exponent $z$ for Model A dynamics, which are summarized in \cref{tab:critical_exponents}.
The static critical exponents in two spatial dimensions can be obtained analytically from Onsager's solution~\cite{Onsager:1943jn}, and with a very high numerical precision in three dimensions using the conformal bootstrap method~\cite{Kos:2016ysd,Komargodski:2016auf}.
The dynamic critical exponent $z$ of the dissipative Model A, has been determined using a variety of methods, including perturbative $\epsilon$-expansion~\cite{Adzhemyan:2021hvo}, non-perturbative renormalization group calculations~\cite{Duclut:2017}, and Monte-Carlo simulations~\cite{Nightingale:2000,Hasenbusch:2020,Schaefer:2022bfm}.
The respective estimates all lie rather close to each other and within the error bars of previous estimates from classical-statistical simulations \cite{Schweitzer:2020noq,Sieke:2024dns}.
Here, we decided to use the literature results for $z$ from Monte-Carlo simulations~\cite{Nightingale:2000,Hasenbusch:2020}. These have previously been used in the trans-critical protocols of Ref.~\cite{Sieke:2024dns} as well, where they have proven to lead to a successful description of universal non-equilibrium scaling functions for the cumulants of the order parameter up to the fourth order.

\begin{table}[tb]
    \begin{minipage}{0.45\linewidth}
        \centering
        \begin{tabular}{lll}
            \toprule
                  & $d=2$        & $d=3$         \\
            \midrule
            $T_c$ & $4.4629(10)$ & $9.37074(28)$ \\
            $B$   & $2.0203(16)$ & $1.937(17)$   \\
            $B^c$ & $1.7425(13)$ & $1.5291(22)$  \\
            $R_0$ & $3040(410)$  & $110(26)$     \\
            \bottomrule
        \end{tabular}
        \caption{Non-universal critical amplitudes of the model in $d=2$ and $d=3$ spatial dimensions.
            The value of $R_0$ is taken from~\cite{Sieke:2024dns}, all other values have been determined in~\cite{Schweitzer:2020noq}.}
        \label{tab:critical_amplitudes}
    \end{minipage}
    \hfill
    \begin{minipage}{0.45\linewidth}
        \centering
        \begin{tabular}{l S[table-format=2.6] S[table-format=1.8]}
            \toprule
                     & {$d=2$}           & {$d=3$}               \\
            \midrule
            $\beta$  & 0.125             & 0.326419 \pm 0.000003 \\
            $\delta$ & 15.0              & 4.78984 \pm 0.00001   \\
            $\nu$    & 1.0               & 0.629971 \pm 0.000004 \\
            $\omega$ & 2.0               & 0.82966 \pm 0.00009   \\
            $z$      & 2.1667 \pm 0.0005 & 2.0245 \pm 0.0015     \\
            \bottomrule
        \end{tabular}
        \caption{Literature values for the static and dynamic critical exponents of the $d=2$ and $d=3$ Ising universality classes, with Model A dynamics.
            For $d=2$ the static critical exponents are known analytically from Onsager's solution~\protect{\cite{Onsager:1943jn}}. For $d=3$ they were determined to high precision via the conformal bootstrap method~\protect{\cite{Kos:2016ysd,Komargodski:2016auf}}.
            For the dynamic critical exponent $z$ we use the values
            from Monte-Carlo simulations~\protect{\cite{Nightingale:2000,Hasenbusch:2020}}.
        }
        \label{tab:critical_exponents}
    \end{minipage}
\end{table}

The scaling ansatz in \cref{eq:scaling_ansatz} allows to define the universal scaling functions by appropriately  choosing the rescaling factor $b$ such that one of the three arguments of the magnetization is transformed to unity, leading to the three different but related two-variable scaling functions $f_\tau$, $f_J$, and $f_R$ as follows:
\begin{alignat}{3}
     & \bar M(\tau, J, R)           &  & = (-\tau)^{\beta} \,  f_\tau\left(\bar J / (-\tau)^{\beta\delta}, \bar R / (-\tau)^{\nu r_c}\right)            &  & \equiv (-\tau)^\beta \, f_\tau(x_\tau, y_\tau), \label{eq:tau_scaling} \\
     & \phantom{\bar M(\tau, J, R)} &  & =  \bar J^{1/\delta}\,  f_J\left(-\tau /\bar J^{1/\beta\delta}, \bar R /\bar J^{\nu r_c/\beta\delta}\right)    &  & \equiv \bar J^{1/\delta} \, f_J(x_J, y_J), \label{eq:J_scaling}        \\
     & \phantom{\bar M(\tau, J, R)} &  & =  \bar R^{\beta/\nu r_c}\,  f_R\left(-\tau / \bar R^{1/\nu r_c}, \bar J / \bar R^{\beta\delta/\nu r_c}\right) &  & \equiv \bar R^{\beta/\nu r_c} \, f_R(x_R, y_R), \label{eq:fts}
\end{alignat}
where the non-universal amplitudes have been defined in such a way that the normalizations $f_\tau(0,0) = f_J(0,0) = f_R(0,0) = 1$ are ensured.
All three scaling functions contain the same information and are related to each other via appropriate rescalings of their arguments such that any one of them can be obtained from any one of the others,
e.g.
\begin{align}
    f_R(x_R, y_R) & = x_R^{\beta} \, f_\tau\left(y_R / x_R^{\beta\delta}, 1/x_R^{\nu r_c}\right).
    \label{eq:fR_from_fTau}
\end{align}
We will focus in particular on the two forms presented in \cref{eq:tau_scaling,eq:fts} as these intuitively describe the deviations from equilibrium scaling (here obtained with $f_\tau(x_\tau,0)$) due to the finite quench rate $R$, and the deviations from the trans-critical finite-time scaling (here obtained with $f_R(0, y_R)$)
due to the non-vanishing reduced temperature $\tau <0$ in the near-critical first-order quenches, respectively.

When the quench is sufficiently slow such that $y_\tau = \bar R/(-\tau)^{\nu r_c} \ll 1$ and $y_J = \bar R /\bar J^{\nu r_c/\beta\delta} \ll 1$, the system is able to relax fast enough to follow the equilibrium magnetization quasi-adiabatically in the slowly changing external field.
In this case, dropping the vanishing scaling variables from the list of arguments in the scaling functions to simplify the notation, standard equilibrium scaling behavior is recovered from \cref{eq:tau_scaling,eq:J_scaling} as
\begin{equation}
    \bar M(\tau, J) \to (-\tau)^{\beta}\,  f_\tau(x_\tau) = \bar J^{1/\delta}\, f_J(x_J), \text{ for } \bar R \to 0,
    \label{eq:equilibrium_scaling}
\end{equation}
where $ f_\tau(x_\tau)\equiv f_\tau(x_\tau,0)$ and $ f_J(x_J) \equiv  f_J(x_J,0)$.
At the same time, with $\bar R \to 0$, the relation in \cref{eq:fR_from_fTau} yields the asymptotic behavior of the finite-time scaling function $f_R$ for large $x_R$ and $y_R \sim x_R^{\beta\delta} $ as
\begin{equation}
    f_R(x_R, y_R) \to x_R^{\beta}\, f_\tau(y_R/x_R^{\beta\delta}), \text{ for } x_R \to \infty.
    \label{eq:fts_to_equilibrium}
\end{equation}

In quenches across the critical point at $\tau = 0$, as previously investigated in~\cite{Sieke:2024dns}, Kibble-Zurek or finite-time scaling results in the magnetization from \cref{eq:J_scaling,eq:fts} with $x_J=x_R=0 $, here denoting $ f_J\left(y_J\right) \equiv f_J\left(0, y_J\right)$ and $f_R\left(y_R \right) \equiv f_R\left(0, y_R \right)$, behaving as
\begin{equation}
    \bar M(J, R) = \bar J ^{1/\delta} f_J\left(y_J\right) = \bar R^{\beta/\nu r_c} f_R\left(y_R \right) .
\end{equation}

In quenches across the first-order transition line, for $\tau < 0$, we therefore expect finite-time scaling to still be observable if  $x_J = -\tau / \bar J^{1/\beta\delta} \ll 1 $ and $x_R = -\tau / \bar R^{1/\nu r_c} \ll 1$, i.e.\ if the quench is sufficiently fast and the driving field is sufficiently strong,
\begin{equation}
    \bar R \gg (-\tau)^{\nu r_c}, \quad \bar J \gg (-\tau)^{\beta\delta}.
\end{equation}
It is important to remember that if any of the parameters $\tau$, $\bar J$ or $\bar R$ get too large, corrections to the scaling behavior in \cref{eq:tau_scaling,eq:J_scaling,eq:fts} will eventually  become relevant.
As seen from the relations above, however, if these corrections turn out to be small, universal finite-time scaling behavior may still be observable well below the critical temperature in sufficiently rapidly driven phase transitions.

To investigate this, we will focus in particular on two observables that characterize the hysteresis behavior of the magnetization curve:
The first is the \emph{remanent magnetization} $M_r \equiv M(J=0)$, i.e.\ the out-of-equilibrium magnetization at the moment during the quench when the time dependent external field $J(t)$ vanishes, which we defined to be at $t=0$.
The second one is the \emph{coercive field} $J_c \equiv J(M=0)$ corresponding to the value of the external field $J$ at the time $t>0$ when the magnetization changes sign.

According to the scaling forms in \cref{eq:tau_scaling,eq:fts}, we expect the remanent magnetization to scale as
\begin{align}
    \bar M_r(\tau, R) = (-\tau)^\beta f_\tau\left(y_\tau\right) = \bar R^{\beta/\nu r_c} f_R\left(x_R\right),
    \label{eq:Mr_scaling}
\end{align}
while the coercive field probes the zeros of the corresponding two-variable scaling functions,
\begin{align}
    0 & = f_\tau\left(x_\tau, y_\tau\right)|_{J=J_c}
    = f_R\left(x_R, y_R\right)|_{J=J_c}.
\end{align}
As $y_\tau$ and $x_R$ are independent of $J$, the unique zeros of the scaling functions $f_\tau$ and $f_R$ are described by functions $x_\tau(y_\tau)$ and $y_R(x_R)$, respectively, which leads to the following scaling laws for the coercive field:
\begin{alignat}{3}
     & \bar J_c(\tau, R) /(-\tau)^{\beta\delta}             = x_\tau\left(y_\tau\right) \quad &  & \Rightarrow \quad  \bar J_c(\tau, R) &  & = (-\tau)^{\beta\delta} x_\tau\left(y_\tau\right)\label{eq:Jc_crit_scaling_form} , \\
     & \bar J_c(\tau, R) / \bar R^{\beta\delta/\nu r_c}  = y_R\left(x_R\right) \quad          &  & \Rightarrow \quad  \bar J_c(\tau, R) &  & = \bar R^{\beta\delta/\nu r_c} y_R\left(x_R\right).
    \label{eq:Jc_fts_scaling_form}
\end{alignat}

\subsection{Far from the critical point: Nucleation and growth}
At temperatures well below the critical temperature, when relaxation times are far shorter than the external driving timescale, the system generally stays in equilibrium and is able to follow the changes of the external field instantaneously up until the coexistence line at $J=0$ is reached.
There, the initial phase becomes metastable and persists for a finite time before decaying to the new stable phase leading to characteristic hysteresis behavior.
According to standard droplet theory~\cite{Rikvold:1994za,Sides:1999}, the decay of the metastable phase to the stable one may proceed in qualitatively different ways, depending on the hierarchy of three relevant length-scales in the system.
Namely, the system size $L$, the critical droplet size $\mathcal{R}_c$, and the average distance  $\mathcal{R}_0$ that a supercritical droplet can grow before meeting another droplet.

If $L \gg \mathcal{R}_0 \gg \mathcal{R}_c$, many droplets of the stable phase nucleate continuously with droplets smaller than the critical size $\mathcal{R}_c$ decaying and droplets larger than $\mathcal{R}_c$ growing until the system has fully transitioned to the new phase.
In this so-called \emph{multi-droplet regime}~\cite{Rikvold:1994za}, the volume fraction $Y(t)$ of the stable phase at time $t$ is described by Avrami's growth law~\cite{Avrami:1939,Avrami:1940,Avrami:1941,Kolmogorov:1937}
\begin{equation}
    Y(t) = 1 - \exp\left[- \int_0^t I(T,J)
        V(t,t_n) dt_n\right], \quad \text{with } V(t,t_n) = \Omega_d \left[\int_{t_n}^t v(t') dt'\right]^d,
\end{equation}
where $V(t,t_n)$ is the volume of a droplet nucleated at time $t_n$ which has grown until time $t$ with a growth velocity $v(t)$, and $I(T,J)$ is the nucleation rate\cite{Rikvold:1994za,Sides:1999,Zhong:2018}.
The coercive field $J_c$ where $M=0$ can then be estimated by solving for the time $t_c$ when the volume fraction of the stable phase reaches $Y(t_c) = 1/2$.
This however, may require further approximations, e.g. the Lifshitz-Allen-Cahn approximation~\cite{Gunton:1983,Lifshitz:1962,Allen:1979} which assumes a linear dependence of the droplet growth velocity on the external field, i.e. $v(t) \propto J(t)$, and an adiabatic approximation for the time-dependence of the nucleation rate~\cite{Zhong:2018}, by which the constant field $J$ in the nucleation rate is replaced with the time-dependent one $J(t)=Rt$.
The resulting implicit equation for the coercive field was found to be in good agreement with numerical simulations of the $d=2$ Ising model in the case of linear driving~\cite{Zhong:2018} as well as in a sinusoidally oscillating field~\cite{Sides:1999}.
However, the authors of \cite{Zhong:2018} also showed that in this multi-droplet regime, no simple scaling behavior of the coercive field with quench rate of the form $J_c - J_{c_0} \propto R^{\alpha}$ with constants $J_{c_0}$ and $\alpha$ is observed.
Recently, scaling behavior in terms of the scaling variable $J(t)(\ln t)^{\kappa}$ was proposed for magnetic first-order phase transitions in Ising systems~\cite{Pelissetto:2025pzi}. However, some discrepancies between the predicted and observed value of the exponent $\kappa$ in three dimensions are not fully understood yet, which is why we will investigate this scaling behavior in more detail in \ref{sec:spinodal_like_scaling}.

In smaller systems, such that $\mathcal{R}_0 \gg L \gg \mathcal{R}_c$, the system changes phase via random nucleation of a single critical droplet, which subsequently grows to fill the entire system before another droplet can nucleate.
This regime is therefore referred to as the \emph{single-droplet regime}.
In this case, the time for the system to transition is dominated by the average nucleation time of the single critical droplet, allowing for the growth-time to be neglected.
The leading asymptotic behavior of the coercive field for very slow quench rates $R \to 0$ was argued to be~\cite{Zhong:2018,Thomas:1993}
\begin{equation}
    J_c \sim (-\ln R)^{-1/(d-1)}.
    \label{eq:single_droplet_scaling}
\end{equation}
However, previous works indicated that the asymptotic logarithmic behavior may only set in at extremely slow quench rates, which are difficult to access in numerical simulations~\cite{Sides:1998,Sides:1999,Zhong:2018}.

When the system size is even smaller than the critical droplet size, i.e. $\mathcal{R}_0 \gg \mathcal{R}_c \gg L$, the behavior of the system is similar to that on the coexistence line $J=0$.
Thermal fluctuations dominate in this \emph{coexistence regime} and the system transitions via spontaneous spin flips of the entire system.
As this behavior is only observed in very small systems and weak fields, it is not relevant for this work.

Finally, for very strong fields, which can be reached in rapid quenches, when the timescale for nucleation to occur is significantly longer than the timescale imposed by the quench rate, the nucleation picture can no longer adequately describe the phase transition, as the system decays via long-wavelength instabilities, similar to spinodal decomposition~\cite{Rikvold:1994za,Gunton:1983}.
The crossover from the multi-droplet regime to this \emph{strong-field regime} occurs at the mean-field spinodal.
Mean-field theory may then describe the relaxation dynamics in this regime, as long as spatial correlations remain sufficiently short-ranged and the system can be considered approximately spatially uniform~\cite{Tomita:1992}.

\subsection{Mean-field theory}
\label{sec:mean_field_theory}
In the mean-field approximation, where fluctuations are neglected, the order-parameter field $\phi(\vec{x})$ takes on the spatially uniform value $M$, and the free energy density can be written as
\begin{equation}
    \frac{F(M)}{V} = \frac{m^2}{2} M^2 + \frac{\lambda}{4!}M^4 - J M.
    \label{eq:mf_free_energy}
\end{equation}
The equilibrium equation of state can be determined via minimization of \cref{eq:mf_free_energy}, which yields
\begin{equation}
    J(M) = M\left(m^2 + \frac{\lambda}{3!}M^2\right).
\end{equation}
Due to local stability of the free energy, the phase transition can only occur beyond a spinodal point $\left|M\right| \geq \left|M_s\right|$,
\begin{equation}
    M_s = \pm\sqrt{-\frac{2m^2}{\lambda}},\quad J_s  =
    \frac{2 m^2}{3} M_s,
    \label{eq:mean_field_spinodal}
\end{equation}
where the free-energy barrier between the two ordered phases vanishes.
To study first-order phase transitions, it is therefore reasonable to shift the order parameter by the spinodal value, and define
\begin{equation}
    \tilde M = M - M_s.
\end{equation}
In this case, the LGW free energy density can be written as
\begin{equation}
    \frac{F(\tilde M)}{V} = \frac{F(M_s)}{V} + \frac{\tilde{m}^2}{2} \tilde M^2 + \frac{v}{3!}\tilde M^3+\frac{\lambda}{4!}\tilde M^4 - \tilde{J}\tilde M,
    \label{eq:free_energy_shifted}
\end{equation}
with
\begin{align}
    \tilde{m}^2 & = m^2 + \frac{\lambda}{2}M_s^2,                                 \\
    v           & = \lambda M_s,                                                  \\
    \tilde{J}   & = J - M_s \left(m^2 + \frac{\lambda}{3!}M_s^2\right) = J - J_s.
\end{align}
When the shifted order parameter $\tilde M$ is small, i.e. near the spinodal point, the cubic term in the free energy will dominate over the quartic one, and the dynamics become governed by an effective cubic theory~\cite{Zhang:2025_CPL}.
The spinodal point itself then corresponds to a critical point of the cubic theory with associated critical exponents that turn out to be equivalent to those of the Yang-Lee edge singularity~\cite{Fisher:1978,Zhong:2012tb,Zhong:2025_CPL}.

We will start our investigation of driven first-order phase transitions in the next section from the mean-field approximation by numerically solving the mean-field equation of motion
\begin{equation}
    \frac{d^2 M}{d t^2} + \gamma \frac{d M}{d t} = -\left(m^2 M + \frac{\lambda}{3!}M^3 - J(t)\right), \quad\text{with } J(t) = R t.
    \label{eq:mean_field_eom}
\end{equation}
Varying the quench rate $R$ allows us to probe the dynamic behavior of the system for a wide range of imposed timescales.
If finite-time scaling holds, we expect the magnetization in sufficiently slow quenches, so that the transition occurs near the spinodal point, to behave according to
\begin{equation}
    M - M_s = R^{\beta/\nu r_c} f_{\tilde M}\left((J-J_s) R^{-\beta\delta/\nu r_c}\right),
    \label{eq:spinodal_fts}
\end{equation}
with $\phi^3$ mean-field exponents $\beta=1$, $\delta=2$, $\nu=\frac{1}{2}$~\cite{Zhong:2012tb,Zhong:2024baf} and $r_c = z + \beta\delta/\nu$.
In fast quenches, when the system falls out of equilibrium early and the transition occurs far beyond the spinodal point, the standard finite-time scaling behavior
\begin{equation}
    M = R^{\beta/\nu r_c} f_M\left(J R^{-\beta\delta/\nu r_c}\right)
    \label{eq:mean_field_fts}
\end{equation}
with $\phi^4$ mean-field exponents $\beta=\frac{1}{2}$, $\delta=3$, $\nu=\frac{1}{2}$~\cite{Zhong:2012tb,Zhong:2024baf} should be recovered.

The value of the dynamic exponent $z$ depends on whether the inertial $\frac{d^2 M}{dt^2}$ term in the equation of motion influences the dynamics or not.
In the overdamped limit $\gamma \left|\frac{\partial M}{\partial t}\right| \gg \left|\frac{\partial^2 M}{\partial t^2}\right|$, when the inertial term is negligible, one finds ${z=2}$~\cite{Kibble:1976sj,Puebla:2018,delCampo:2013nla,Laguna:1997}.
If however, the inertial term is relevant, i.e. when the system is probed on timescales shorter than the typical damping time $\gamma^{-1}$, which is the case for sufficiently fast quenches, one finds $z=1$~\cite{Puebla:2018,delCampo:2013nla,Laguna:1998,Moro:1999,Chiara:2010}.
We therefore expect a crossover from $z=2$ to $z=1$ as the quench rate is increased, and consequently expect to find hysteresis exponents $\beta/\nu r_c = 1/3$ and $\beta\delta/\nu r_c = 3/4$ in asymptotically fast quenches, instead of the typically reported values $\beta/\nu r_c = 1/5$ and $\beta\delta/\nu r_c = 3/5$ that apply to the overdamped Langevin equations in mean-field approximation~\cite{Zhong:2012tb}.

\section{Mean-field results}
\label{sec:mean_field_results}

\begin{figure}[tb]
    \includegraphics[width=.498\textwidth]{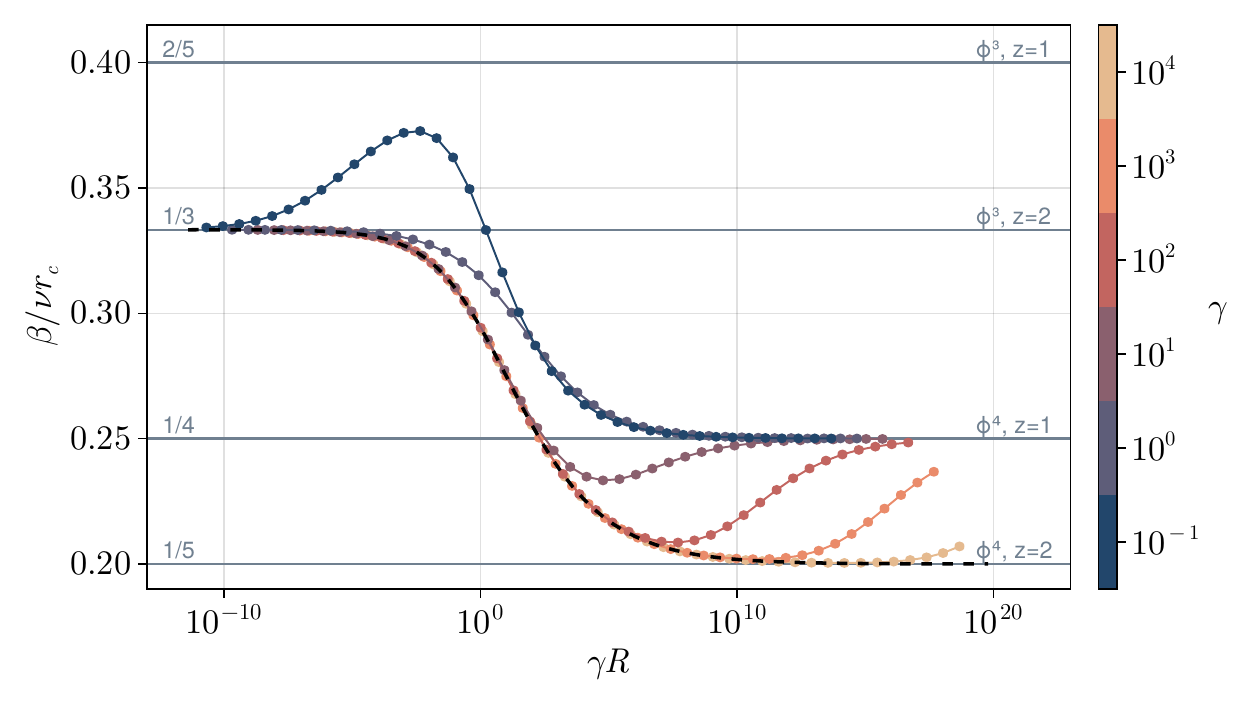}
    \includegraphics[width=.498\textwidth]{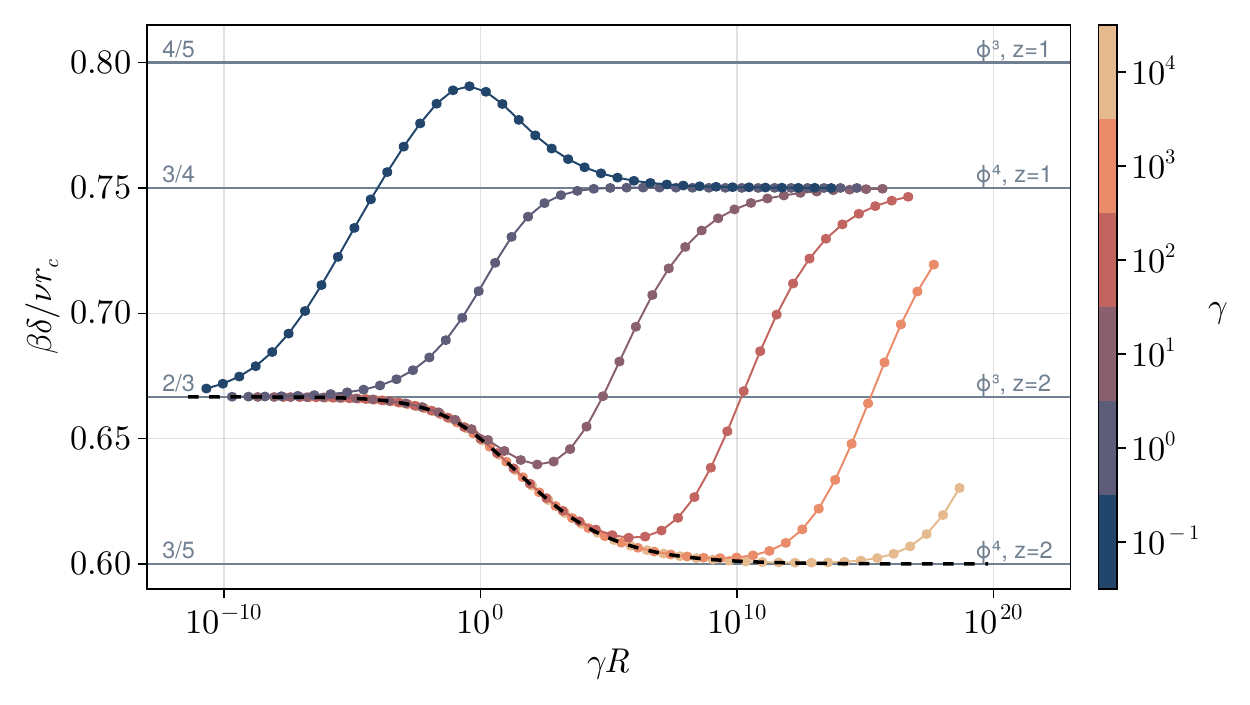}
    \caption{
        \label{fig:mean_field_exponents}
        Quench rate dependence of the mean-field hysteresis exponents $\beta/\nu r_c$ (left) and $\beta\delta/\nu r_c$ (right) extracted from the out-of-equilibrium scaling behavior of $M(J_s) - M_s$ and $J(M_s) - J_s$ respectively.
        The black dashed line shows the result in the overdamped limit.
        The colored data points show results for different values of the damping coefficient $\gamma$.
        Lines connecting the data points serve only as a guide to the eye.}
\end{figure}

To verify the expectations set in the previous section, we have numerically integrated the mean-field equation of motion \cref{eq:mean_field_eom} for different quench rates $R$ and investigated the scaling behavior of the characteristic points $M(J=J_s)$ and $J(M=M_s)$.
According to the scaling hypothesis \cref{eq:spinodal_fts} we expect
\begin{align}
    M(J=J_s) - M_s & \sim R^{\beta/\nu r_c},       \\
    J(M=M_s) - J_s & \sim R^{\beta\delta/\nu r_c},
\end{align}
which allows us to extract the hysteresis exponents $\beta/\nu r_c$ and $\beta\delta/\nu r_c$ associated with the magnetization and the external field respectively, by computing numerical derivatives of the corresponding double-logarithmic data.
The results of this analysis are shown in \cref{fig:mean_field_exponents} where we present the numerically obtained effective hysteresis exponents as functions of the quench rate for different values of the damping coefficient $\gamma$.
Since $\gamma$ only defines the overall timescale of the dynamics in the overdamped limit, we present the results as functions of $\gamma R$ to allow for the convergence of the numerical data to the same overdamped limit for different values of $\gamma$.

\cref{fig:mean_field_exponents} allows us to clearly identify three distinct scaling regimes depending on the quench rate $R$ and the damping coefficient $\gamma$.
In the overdamped limit (black dashed curve), the static critical exponents $\beta, \delta, \nu$ match the expected $\phi^3$ mean-field values in slow quenches, i.e. $R \ll 1/\gamma$, and cross over to $\phi^4$ mean-field values in fast quenches, while the dynamic exponent remains $z=2$ throughout, resulting in the hysteresis exponent $\beta/\nu r_c$ crossing over from $1/3$ to $1/5$, and $\beta\delta/\nu r_c$ crossing over from $2/3$ to $3/5$ as the quench rate is increased.

For finite values of the damping coefficient $\gamma$, we observe that the dynamic critical exponent crosses over from $z=2$ to $z=1$ as the quench rate is increased and the inertial term in the equation of motion gains relevance.
This results in the hysteresis exponents asymptotically approaching $\beta/\nu r_c = 1/4$ and $\beta\delta/\nu r_c = 3/4$ in fast quenches.
For large values of $\gamma \gg 1$, there is an intermediate regime where overdamped behavior with $z=2$ is still observable in fast quenches, before crossing over to the inertial regime with $z=1$ at even larger quench rates.
In asymptotically slow quenches, however, the expected spinodal scaling with $\phi^3$ mean-field exponents and $z=2$ is always approached, independent of the damping strength.

If the damping is sufficiently weak, we expect another scaling regime to arise for intermediate quench rates, where the system transitions close to the mean-field spinodal point, but the inertial term still dominating the dynamics leading to $z=1$, and the static exponents retaining their $\phi^3$ mean-field values.
In this case, the hysteresis exponents are expected to be $\beta/\nu r_c = 2/5$ and $\beta\delta/\nu r_c = 4/5$.
The peak in the numerically obtained effective  exponents for $\gamma=0.1$ in \cref{fig:mean_field_exponents} support this idea. However, to fully resolve this regime, one would need to solve the mean-field equations for even smaller values of $\gamma$ which is numerically challenging due to the strong oscillatory behavior of the solutions.

\begin{figure}[tb]
    \centering
    \includegraphics[width=.498\textwidth]{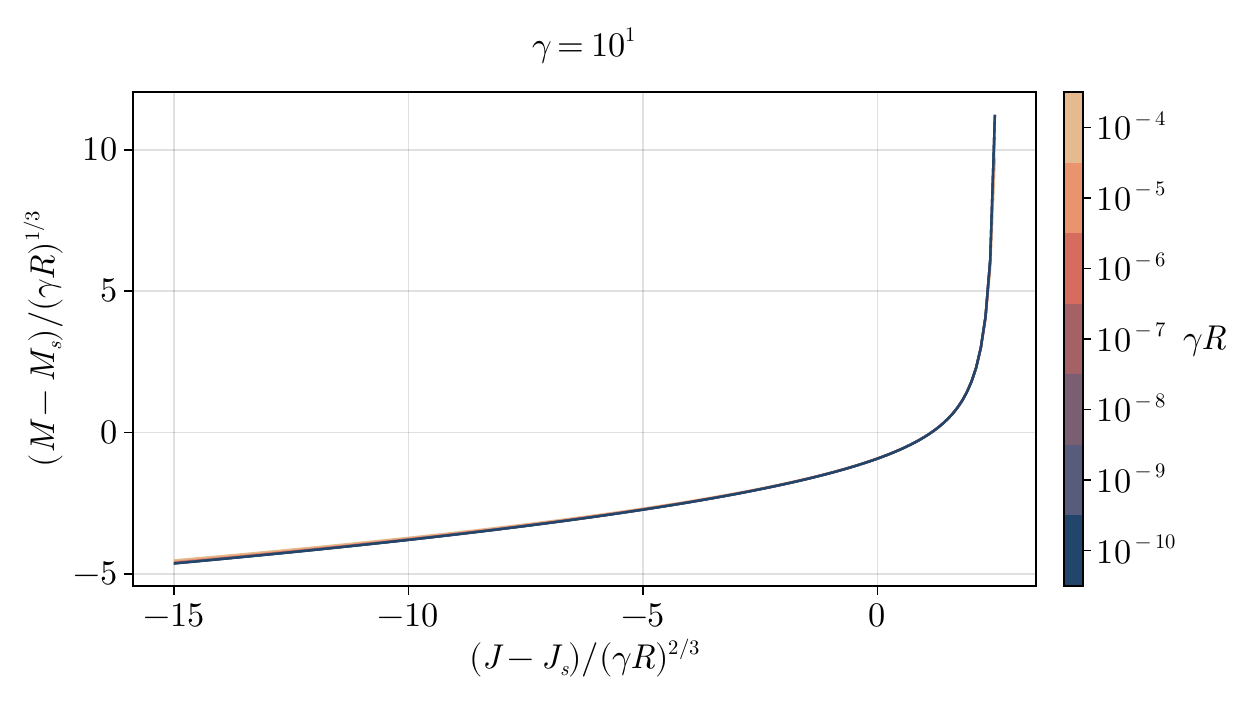}
    \caption{\label{fig:mean_field_scaling_function_spinodal}
        Scaled mean-field magnetization data for multiple slow quench rates $R\ll 1/\gamma$ such that the transition takes place close to the spinodal point. Scaling collapse of the data reveals the mean-field spinodal scaling function $f_{\tilde M}$.
    }
\end{figure}

Having identified the different scaling regimes and numerically verified the hysteresis exponents allows us to compute the mean-field scaling functions $f_{\tilde M}$ and $f_M$ for transitions close to and far from the spinodal point respectively by rescaling the obtained magnetization data according to \cref{eq:spinodal_fts} and \cref{eq:mean_field_fts} in the appropriate regimes.
In \cref{fig:mean_field_scaling_function_spinodal} we present the scaling function $f_{\tilde M}$ describing the behavior of the magnetization in slow quenches where the transition takes place close to the spinodal point.

\begin{figure}[tb]
    \includegraphics[width=.498\textwidth]{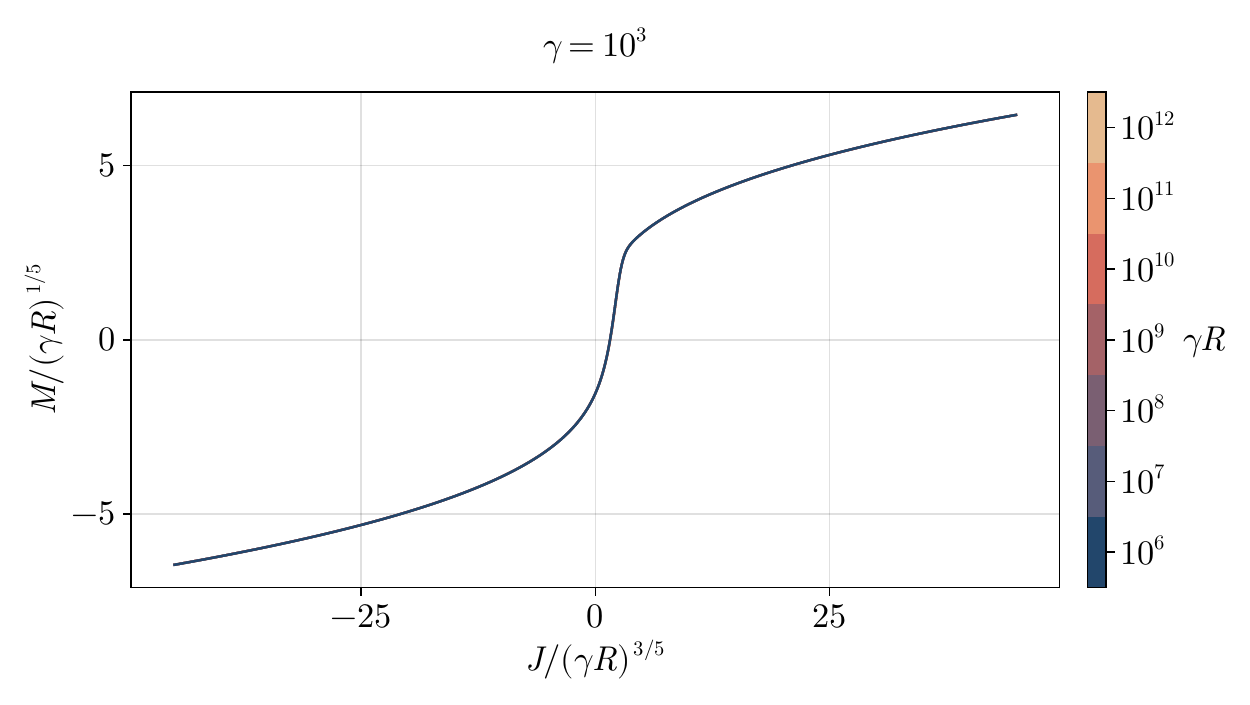}
    \includegraphics[width=.498\textwidth]{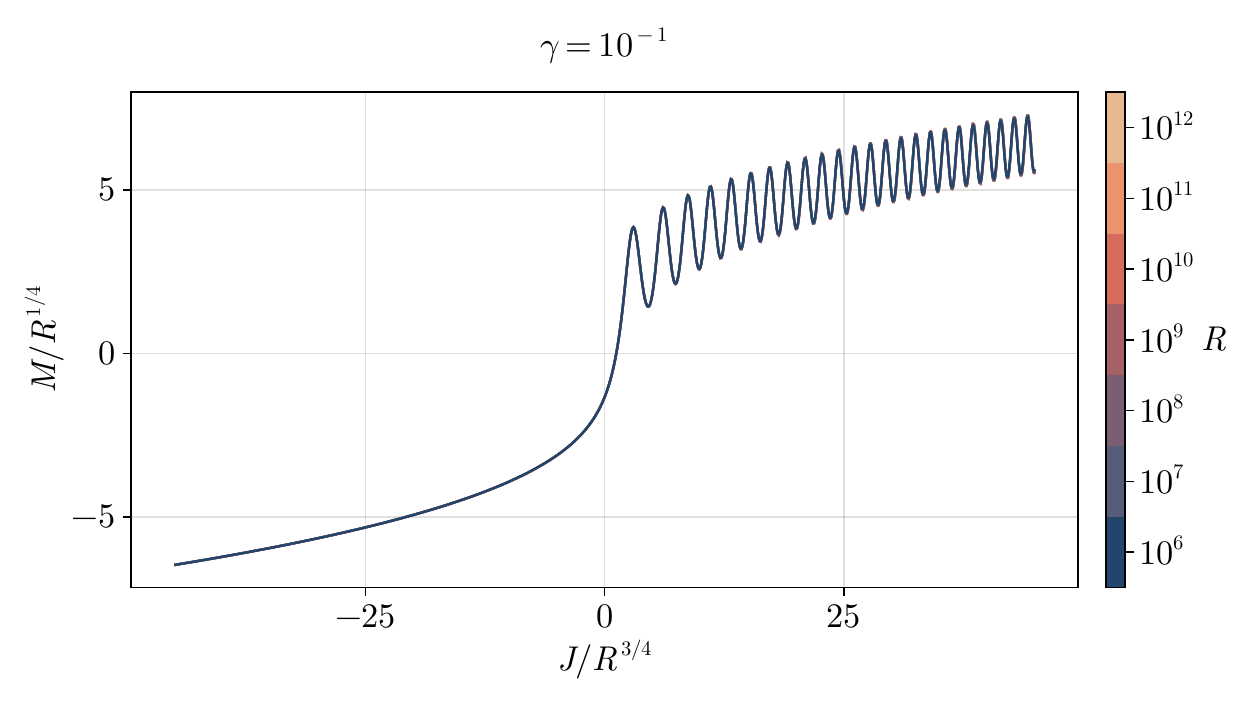}
    \caption{\label{fig:mean_field_scaling_functions} Mean-field scaling function $f_{M}$ extracted from the rescaled magnetization data in fast quenches for different values of the damping coefficient $\gamma$.
        The left panel shows the rescaled magnetization assuming the inertial term is irrelevant, i.e. $\left|\tfrac{d^2 M}{d t^2}\right| \ll \gamma \left|\tfrac{d M}{d t}\right|$, leading to $z=2$ which corresponds to the overdamped limit of the Langevin \cref{eq:mean_field_eom}, while the right panel shows the rescaled magnetization assuming a dominant inertial term, i.e. $\left|\tfrac{d^2 M}{d t^2}\right| \gg \gamma \left|\tfrac{d M}{d t}\right|$, resulting in $z=1$.}
\end{figure}

In faster quenches, the system falls out of equilibrium earlier, and the transition occurs long after the spinodal point is crossed.
The shifted order parameter $\tilde M = M - M_s$ can then no longer be consider small, and the quartic $\tilde M^4$ term in the free energy \cref{eq:free_energy_shifted} becomes relevant again.
In this case, the spinodal scaling function $f_{\tilde M}$ does not adequately describe the dynamic behavior of the magnetization.
Instead, the standard finite-time scaling ansatz \cref{eq:mean_field_fts} applies, and the scaling function $f_M$ can be extracted from the rescaled magnetization data as shown in \cref{fig:mean_field_scaling_functions}.
If the underdamped Langevin equation is considered, for fast enough quench rates, the magnetization will always be described by the form of the scaling function $f_M$ shown on the right-hand side of \cref{fig:mean_field_scaling_functions}
exhibiting dynamic scaling behavior with a dynamic exponent $z=1$.

However, if the damping coefficient $\gamma$ is sufficiently large, there can exist an intermediate regime of quench rates where the overdamped Langevin equation effectively describes the dynamics. In this case, the scaling function $f_M$ shown on the left-hand side of \cref{fig:mean_field_scaling_functions} applies instead, which describes overdamped dynamic non-equilibrium scaling behavior with $z=2$.

Notably, both forms of the scaling function $f_M$ behave visibly identical prior to the transition point where the magnetization changes sign.
Their differences only become apparent after the transition, where in the case of underdamped dynamics the magnetization exhibits persisting oscillations that are captured by the scaling function $f_M$ shown on the right-hand side of \cref{fig:mean_field_scaling_functions}. While in the overdamped case, shown on the left-hand side of \cref{fig:mean_field_scaling_functions}, these oscillations are absent and purely relaxational behavior is observed.

\section{Numerical results including fluctuations}
\label{sec:numerical_results}

To investigate the effects of fluctuations on driven first-order phase transition, we have performed numerical simulations of the Langevin Eq. (\ref{eq:eom}) in the underdamped regime with $\gamma = 0.1$ in $d=2$ and $d=3$ spatial dimensions for a wide range of reduced temperatures $\tau$ and quench rates $R$. Lattice sizes of $N=510$ in $d=2$ and $N=254$ in $d=3$ were used.
These were found to be large enough for the results to be free of finite-size effects in the considered parameter ranges.
All observables were obtained by averaging over independent simulation runs with different seeds for the random noise in the Langevin equation.
The number of independent simulations ranged from around \num{10000} runs for slow quenches close to the critical point, to around \num{10} runs for very fast quenches at low temperatures, where fluctuations are small and the system behaves nearly deterministically.
Statistical uncertainties were estimated via bootstrap resampling.

\begin{figure}[tb]
    \includegraphics[width=.498\textwidth]{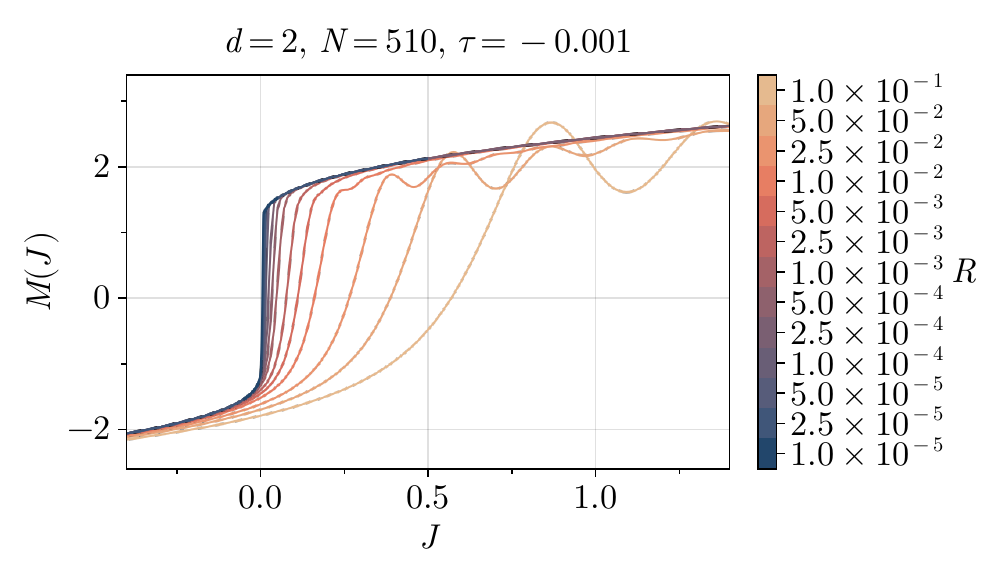}
    \includegraphics[width=.498\textwidth]{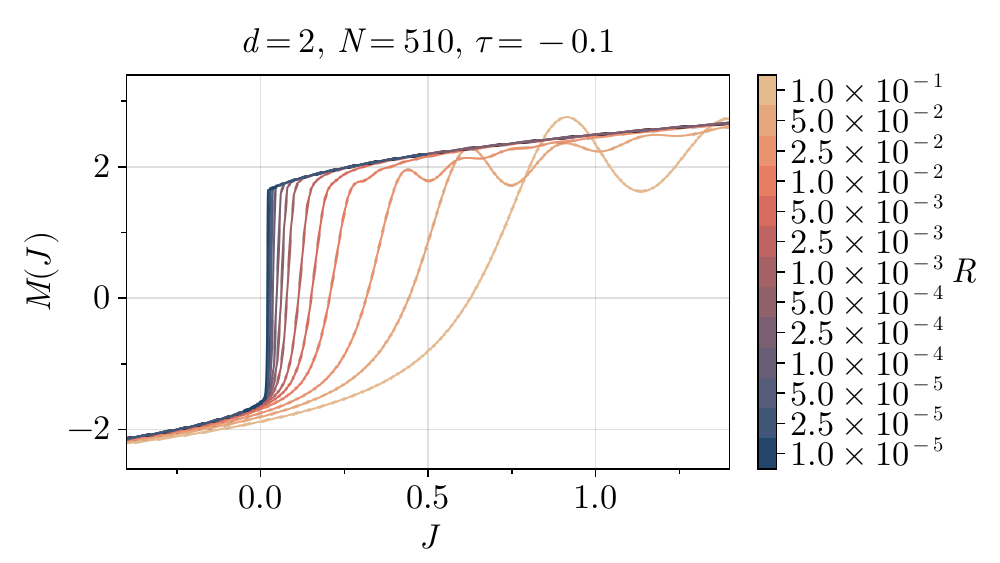}
    \caption{\label{fig:M_vs_J} Magnetization $M$ as a function of external field $J$ for different quench rates $R$ close to the critical temperature ($\tau = -0.001$, left) and away from the critical point ($\tau = -0.1$, right). Statistical uncertainties, drawn as shaded regions, are not visible at this scale.}
\end{figure}

The behavior of the magnetization as a function of the external field is illustrated in \cref{fig:M_vs_J} for different quench rates, where we show selected results for $d=2$ and two different reduced temperatures $\tau$, one close to the critical point ($\tau = -0.001$, left panel) and one further away ($\tau = -0.1$, right panel).
In all cases, we observe hysteresis behavior, with the transition between the two ordered phases occurring at larger values of the external field $J$ in faster quenches.
Comparing the two temperatures, one can see that in slow quenches close to the critical point (left-hand side of \cref{fig:M_vs_J}), the transition occurs close to $J=0$ and appears more rounded, while further away from the critical point (right-hand side of \cref{fig:M_vs_J}) the transition happens away from $J=0$ even in the slowest quenches considered here, and appears sharper.
In fast quenches however, the magnetization curves appear qualitatively similar for the two different temperatures.
Additionally, we observe that in sufficiently fast quenches, the magnetization exhibits oscillatory behavior after the transition, similar to what was observed in the mean-field results presented in \cref{fig:mean_field_scaling_functions}.

To quantify the hysteresis behavior, we have measured the coercive field $J_c \equiv J(M=0)$ as well as the remanent magnetization $M_r \equiv M(J=0)$, as functions of the quench rate $R$ and reduced temperature $\tau$.
We will focus on the results for $J_c$ in the following, whereas the results for $M_r$ will be discussed in the subsequent section.

\subsection{Scaling behavior of the coercive field}
\begin{figure}[tb]
    \includegraphics[width=.498\textwidth]{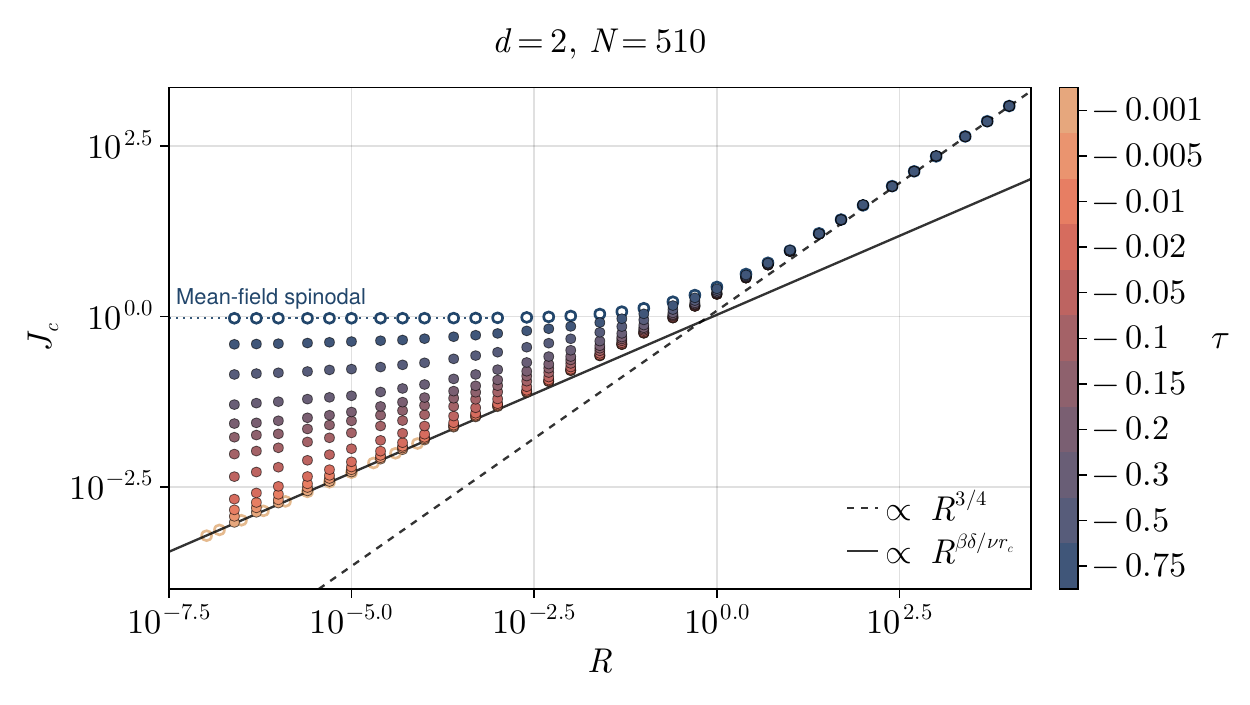}
    \includegraphics[width=.498\textwidth]{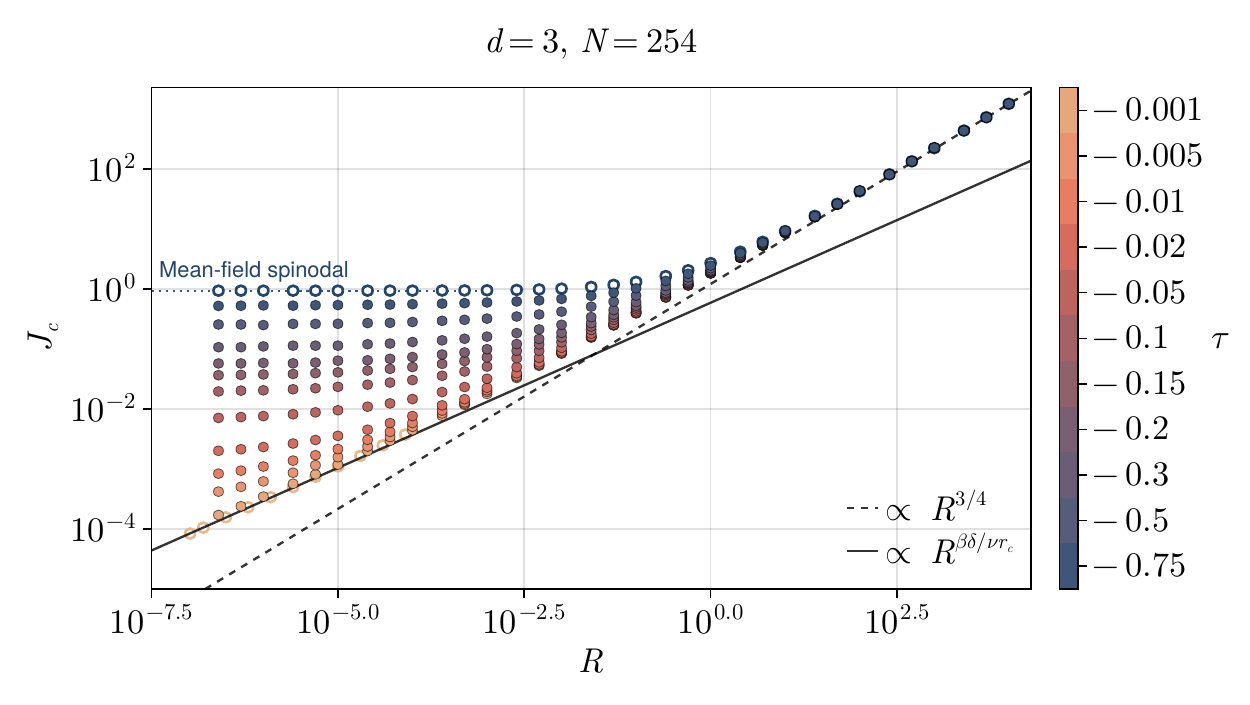}
    \caption{\label{fig:Jc_vs_R} Coercive field $J_c \equiv J(M=0)$ as a function of quench rate $R$ and reduced temperature $\tau$ in $d=2$ (left) and $d=3$ (right) spatial dimensions.
        Open yellow circles correspond to data at the critical temperature $\tau = 0$ obtained in~\cite{Sieke:2024dns}. The expected critical scaling $J_c \sim R^{\beta\delta/\nu r_c}$ is indicated by the solid black line, with the critical exponents being $\beta\delta/\nu r_c \approx 0.464$ for the $d=2$ Ising universality class and $\beta\delta/\nu r_c \approx 0.551$ for $d=3$.
        The precise values of the critical exponents used here, as well as their references are listed in \cref{tab:critical_exponents}.
        Open blue circles show our numerical mean-field results in absence of thermal fluctuations, i.e. $\tau = -1$.
        The theoretical value for the mean-field spinodal is obtained from \cref{eq:mean_field_spinodal} to be $J_s = 2 \sqrt{2} / 3$ and is indicated by the dotted blue line.
        The black dashed line corresponds to mean-field finite-time scaling $J_c \sim R^{3/4}$ for underdamped dynamics with $z=1$.
        Statistical uncertainties of the numerical results including fluctuations were estimated via bootstrap resampling, but are not visible at this scale.
    }
\end{figure}

The measured coercive field as a function of quench rate $R$ and reduced temperature $\tau$ is shown in \cref{fig:Jc_vs_R}, where we also include our mean-field results for comparison (open blue circles),
as well as results for quenches across the critical point $\tau = 0$ which were obtained in~\cite{Sieke:2024dns} (open yellow circles).
We observe qualitatively similar results in two and three spatial dimensions.

In fast quenches $\left(R \gtrsim 1\right)$,
the phase transition occurs beyond the mean-field spinodal indicated by the dotted blue line in \cref{fig:Jc_vs_R}, and the coercive field approaches the mean-field result independently of temperature and number of spatial dimensions.
Furthermore, the coercive field exhibits power-law scaling with the quench rate according to \cref{eq:mean_field_fts}, with an exponent that approaches the mean-field value of $3/4$ expected for underdamped dynamics with a dynamic exponent $z=1$, as discussed in \cref{sec:mean_field_theory}.

In slow quenches $\left(R \lesssim 1\right)$, the mean-field results (open blue circles) asymptotically approach the spinodal value $J_s$, 
exhibiting power-law scaling with $\phi^3$ mean-field exponents and $z=2$, i.e. $J_c - J_s \sim R^{2/3}$, which is not directly visible in \cref{fig:Jc_vs_R} as the spinodal $J_s$  was not subtracted, but this was already illustrated in \cref{fig:mean_field_exponents}.
In contrast, when fluctuations are included, the phase transition can occur before the spinodal point is reached, and one expects the coercive field to eventually vanish in adiabatically slow quenches, i.e. $R \to 0$. However, the limit $J_c \to 0$ is approached only very gradually.
Specifically, according to \cref{eq:single_droplet_scaling}, the asymptotic decay of the coercive field in the single-droplet regime follows a logarithmic dependence on the quench rate.

Approaching the critical temperature in the regime of slow quenches, we observe that standard Kibble-Zurek  scaling $J_c \sim R^{\beta\delta/\nu r_c}$ with critical exponents in agreement with the $d=2$ and $d=3$ Ising universality classes and the dynamic exponent $z$ for Model A dynamics, is recovered as indicated by the solid black lines in \cref{fig:Jc_vs_R} that perfectly match the data at $\tau \approx 0$.
The values of critical exponents used to compute the reference line \cref{fig:Jc_vs_R} are listed in \cref{tab:critical_exponents}.

\begin{figure}[tb]
    \includegraphics[width=.498\textwidth]{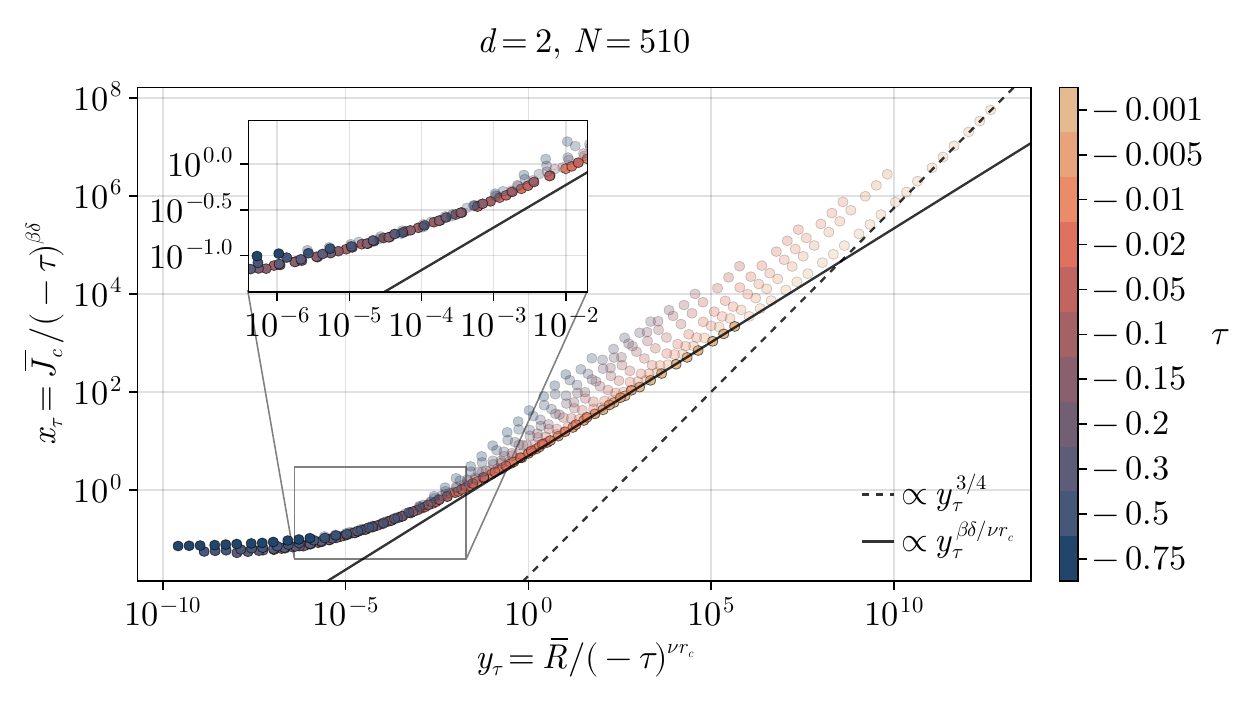}
    \includegraphics[width=.498\textwidth]{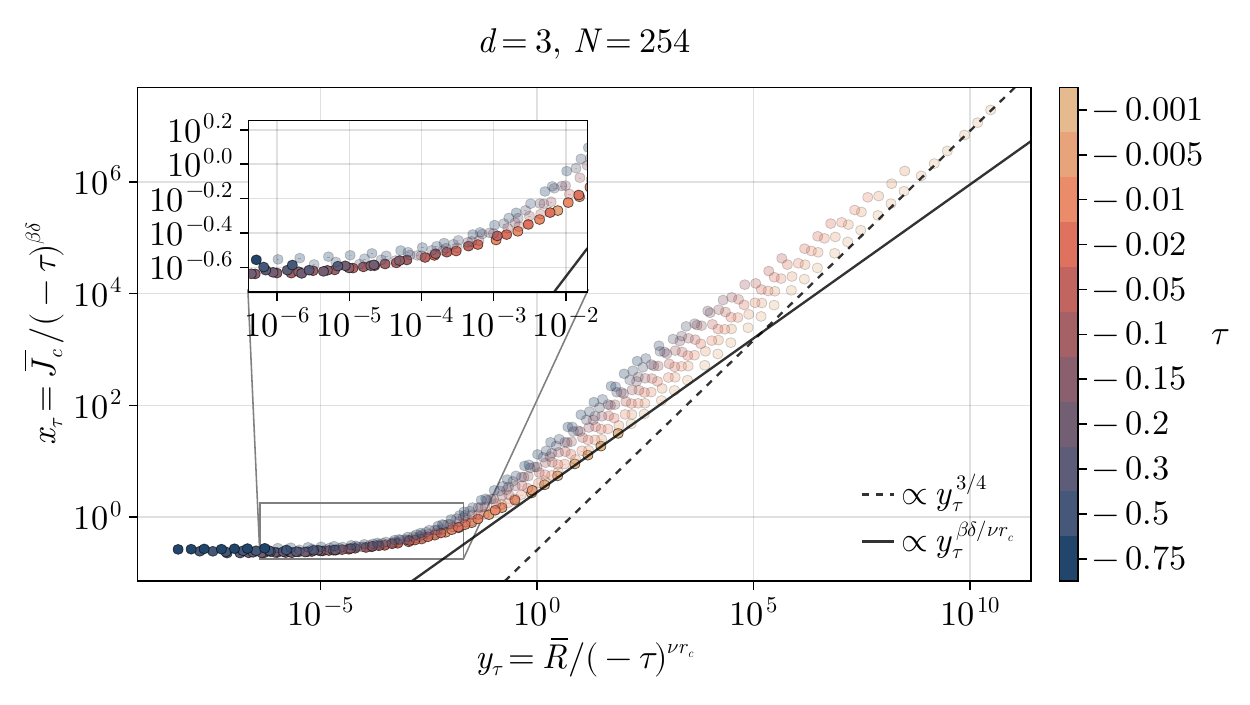}
    \includegraphics[width=.498\textwidth]{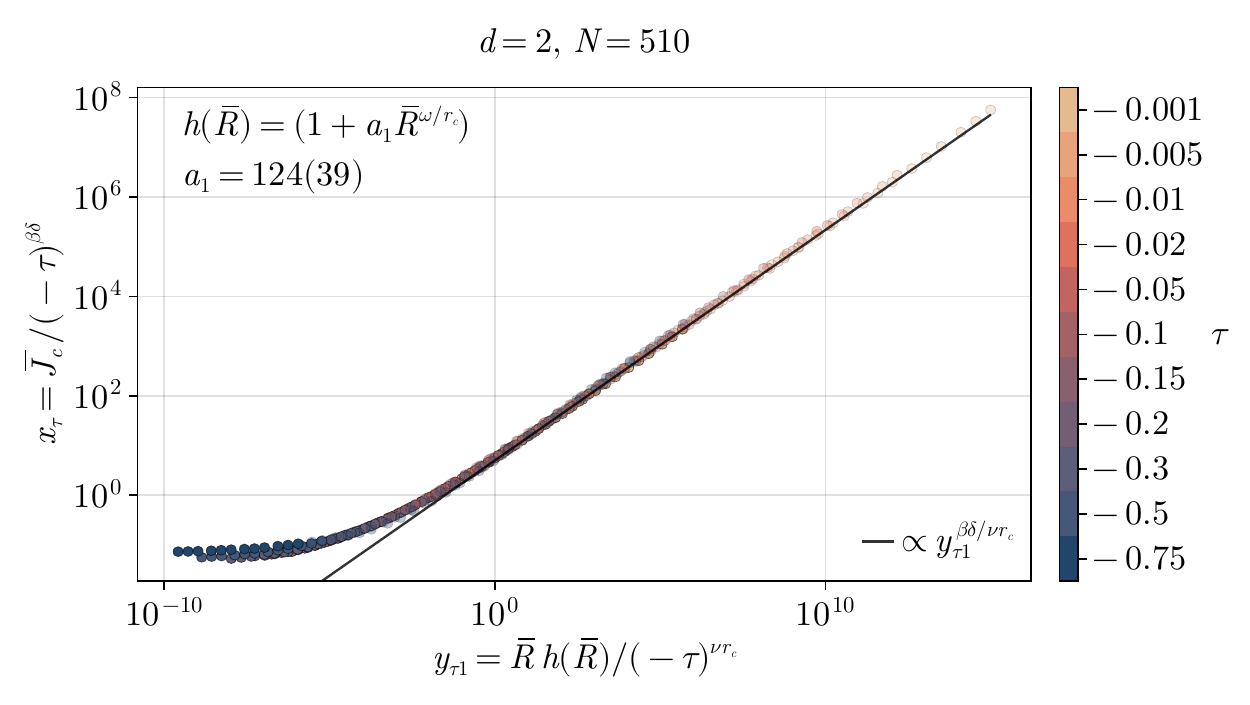}
    \includegraphics[width=.498\textwidth]{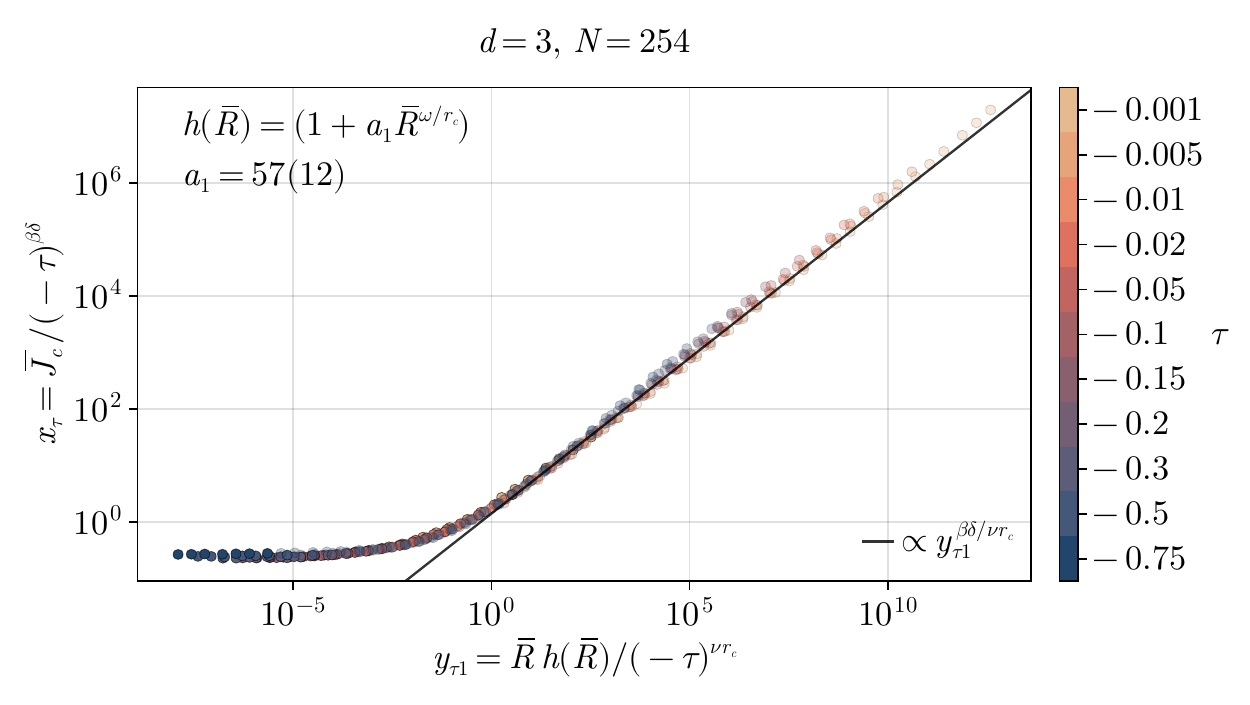}
    \caption{\label{fig:scaled_Jc_vs_R}
        Scaled coercive field in $d=2$ (left) and $d=3$ (right) spatial dimensions. Data points corresponding to fast quenches ($R>10^{-3}$ for $d=2$ and $R>2.5\times 10^{-5}$ for $d=3$) are drawn semi-transparent. The solid black line indicates the expected finite-time scaling behavior with Ising critical exponents.
        The dashed black line indicates finite-time scaling with a mean-field exponent of $3/4$.
        The upper panels show the rescaled data according to \cref{eq:Jc_crit_scaling_form}, while the lower panels show the rescaled data according to the ansatz in \cref{eq:Jc_crit_scaling_form_with_correction} including the leading finite-time scaling correction.}
\end{figure}

By rescaling the coercive field and quench rate with the reduced temperature according to \cref{eq:Jc_crit_scaling_form}, we attempt to collapse the data for different temperatures onto a single universal curve describing the zeros of the universal scaling function $f_\tau(x_\tau,y_\tau)$, defined in \cref{eq:tau_scaling}. The result of this rescaling procedure is shown in the upper panels of \cref{fig:scaled_Jc_vs_R}.

Equivalently, one may also attempt to collapse the data by rescaling the coercive field and reduced temperature with the quench rate according to \cref{eq:Jc_fts_scaling_form}, which correspondingly probes the zeros of the universal function $f_R(x_R,y_R)$ defined in \cref{eq:fts}. This is done in \ref{sec:additional_scaling_functions} where we show in \cref{fig:scaled_Jc_vs_R_fts} the result of this alternative approach containing essentially the same information, but presented in a different way.

In the former approach presented in \cref{fig:scaled_Jc_vs_R}, we find that the data collapse is good for sufficiently slow quenches, namely $R \lesssim 10^{-3}$ in two dimensions and $R\lesssim 2.5\times 10^{-5}$ for the three dimensional model.
The data points corresponding to faster quenches are drawn semi-transparent and are seen to deviate from the critical finite-time scaling behavior, instead crossing over to mean-field scaling indicated by the dashed black line.
We also find that in the equilibrium limit $R \to 0$, the coercive field to leading order behaves as $J_c \sim (-\tau)^{\beta\delta}$, with only small deviations from this behavior observed for the smallest temperatures furthest away from the critical point, where subleading and regular contributions to the scaling behavior are expected to become relevant.
The region where the scaling collapse is found to work best for almost all temperatures considered here is shown in detail in the inset of the upper panels of \cref{fig:scaled_Jc_vs_R}.

The scaling collapse can be significantly improved by including the leading correction to finite-time scaling, which is known to be characterized by $R^{\omega/r_c}$~\cite{Zhong:2011,Liu:2025}, where $\omega$ is the leading correction exponent of the Ising universality class and is also listed in \cref{tab:critical_exponents}.
Considering the finite-time scaling correction to leading order leads to the following modified scaling ansatz for the coercive field
\begin{equation}
    \bar J_c(\tau, R) \approx (-\tau)^{\beta\delta} x_{\tau}\left(\bar R [1 + a_1 \bar R^{\omega/r_c}] / (-\tau)^{\nu r_c}\right),
    \label{eq:Jc_crit_scaling_form_with_correction}
\end{equation}
where $a_1$ is a non-universal constant that is determined by a single fit to the numerical data.

The rescaled data according to the above scaling form in \cref{eq:Jc_crit_scaling_form_with_correction}
as well as the best-fit values for the non-universal constant $a_1$ are shown in the lower panels of \cref{fig:scaled_Jc_vs_R}.
We find that the inclusion of the leading scaling correction significantly improves the data collapse.
Although the data points corresponding to fast quenches seem to perfectly align with the expected Kibble-Zurek scaling behavior shown again as a solid black line in \cref{fig:scaled_Jc_vs_R}, in sufficiently fast quenches, the data points will always deviate from the critical scaling behavior and instead approach mean-field scaling with an exponent of $3/4$. This can be seen more clearly in the alternative scaling collapse shown in \cref{fig:scaled_Jc_vs_R_fts}.
Nevertheless, including the leading finite-time scaling correction appears to significantly extend the dynamic scaling window where the coercive field is described by a universal scaling function.

\subsection{Scaling behavior of the remanent magnetization}
\begin{figure}[tb]
    \includegraphics[width=.498\textwidth]{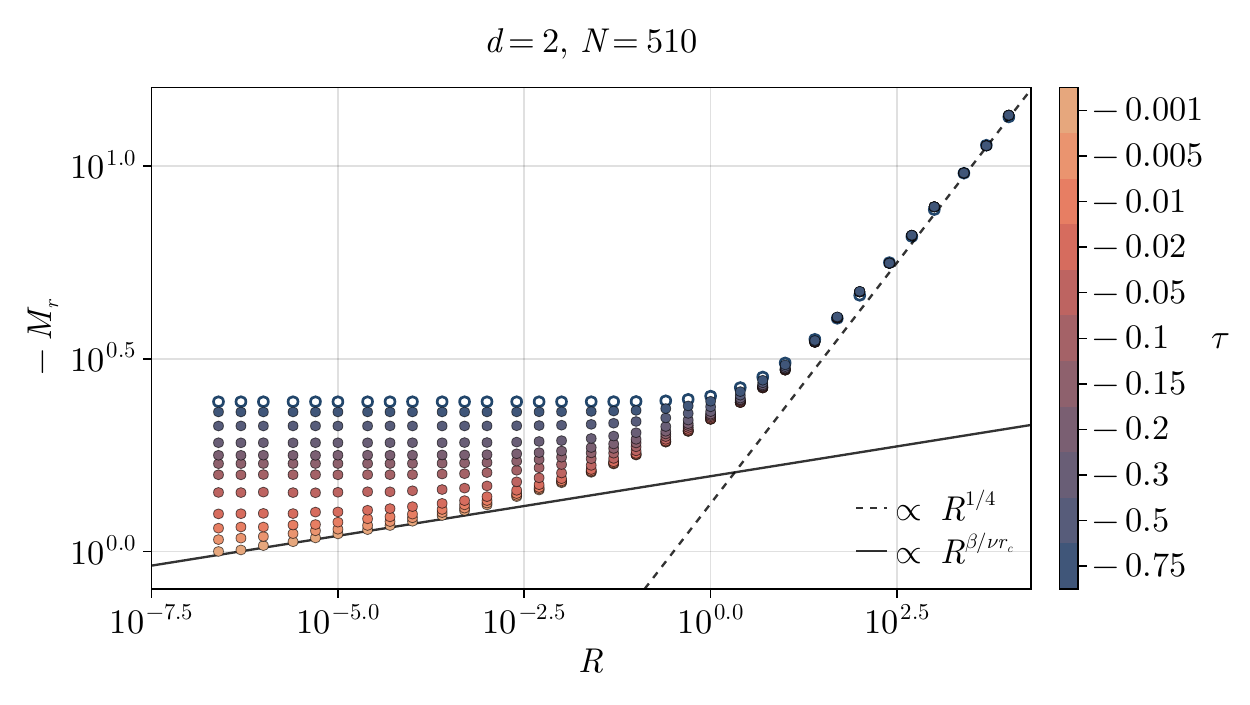}
    \includegraphics[width=.498\textwidth]{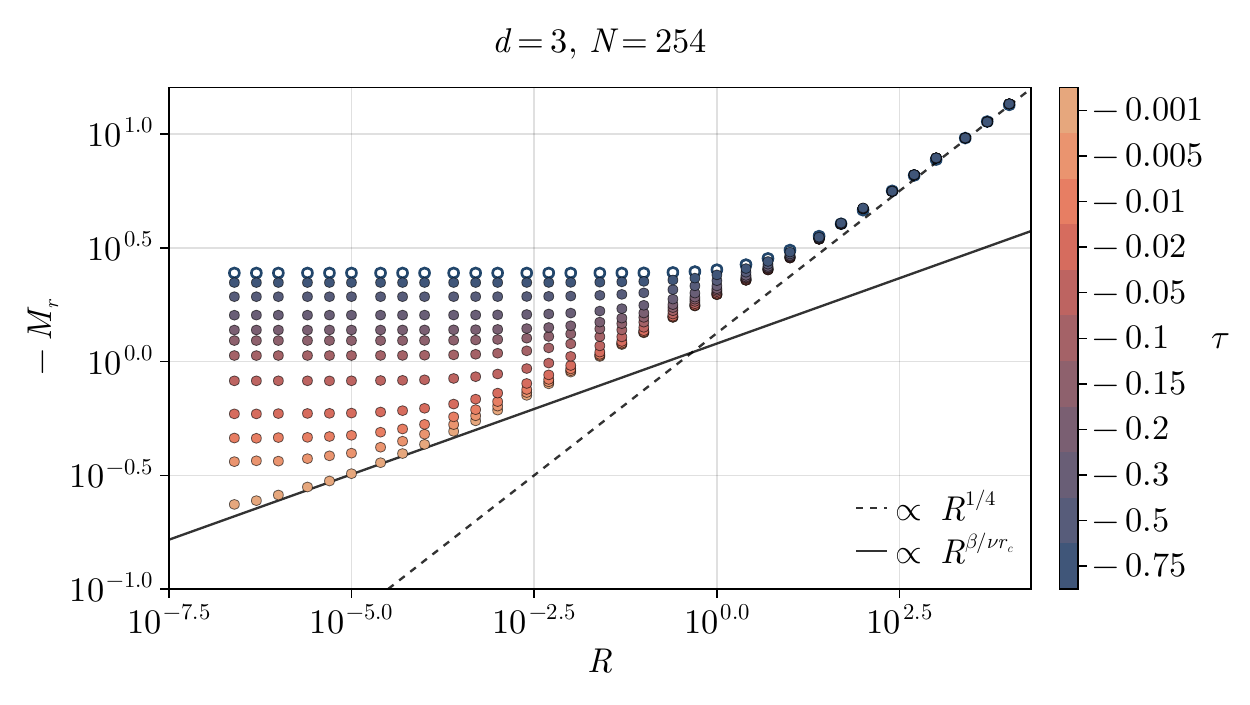}
    \caption{\label{fig:Mr_vs_R}
        Remanent magnetization $M_r \equiv M(J=0)$ as a function of quench rate $R$ and reduced temperature $\tau$ in $d=2$ (left) and $d=3$ (right) spatial dimensions.
        The expected critical scaling $M_r \sim R^{\beta/\nu r_c}$ is indicated by the solid black line, with the exponent being $\beta/\nu r_c \approx 0.031$ for the $d=2$ Ising universality class and $\beta/\nu r_c \approx 0.115$ for $d=3$.
        The values of the critical exponents used, as well as their references are listed in \cref{tab:critical_exponents}.
        Open blue circles again show our numerical mean-field results.
        Statistical uncertainties, estimated via bootstrap resampling, are not visible at this scale.
    }
\end{figure}

A similar analysis can be performed for the remanent magnetization $M_r$ describing the rate-dependent out-of-equilibrium magnetization at vanishing external field, i.e. $M_r \equiv M(J=0)$.
Our numerical results for this observable as a function of quench rate and reduced temperature are presented in \cref{fig:Mr_vs_R}.
As before, we include numerical mean-field results for $M_r$, shown as open blue circles.
We find that, just like the coercive field, the remanent magnetization approaches the mean-field results in sufficiently fast quenches $\left(R \gtrsim 1\right)$ and asymptotically follows underdamped mean-field scaling $M_r \sim R^{1/4}$
indicated by the dashed black line in \cref{fig:Mr_vs_R}.
This is again true, independent of temperature and dimensionality of the system.
However, in contrast to the coercive field, the remanent magnetization $M_r$ does not vanish in the equilibrium limit $R \to 0$, but instead approaches the finite spontaneous magnetization $M_{eq}(\tau)$ which is non-zero for all $\tau < 0$.
Approaching the critical temperature, finite-time scaling behavior emerges with an exponent $\beta/\nu r_c$ in line with expectations for the $d=2$ and $d=3$ Ising universality classes and Model A dynamics, as indicated by the solid black lines in \cref{fig:Mr_vs_R}.

\begin{figure}[tb]
    \includegraphics[width=.498\textwidth]{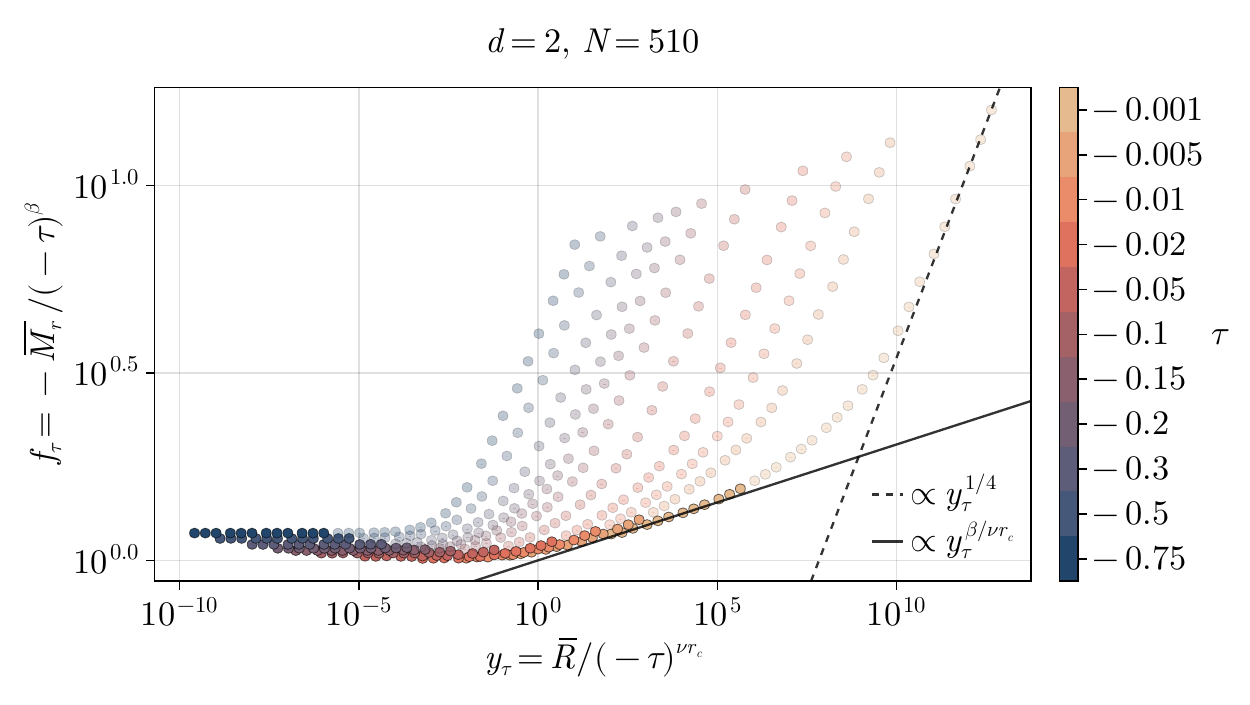}
    \includegraphics[width=.498\textwidth]{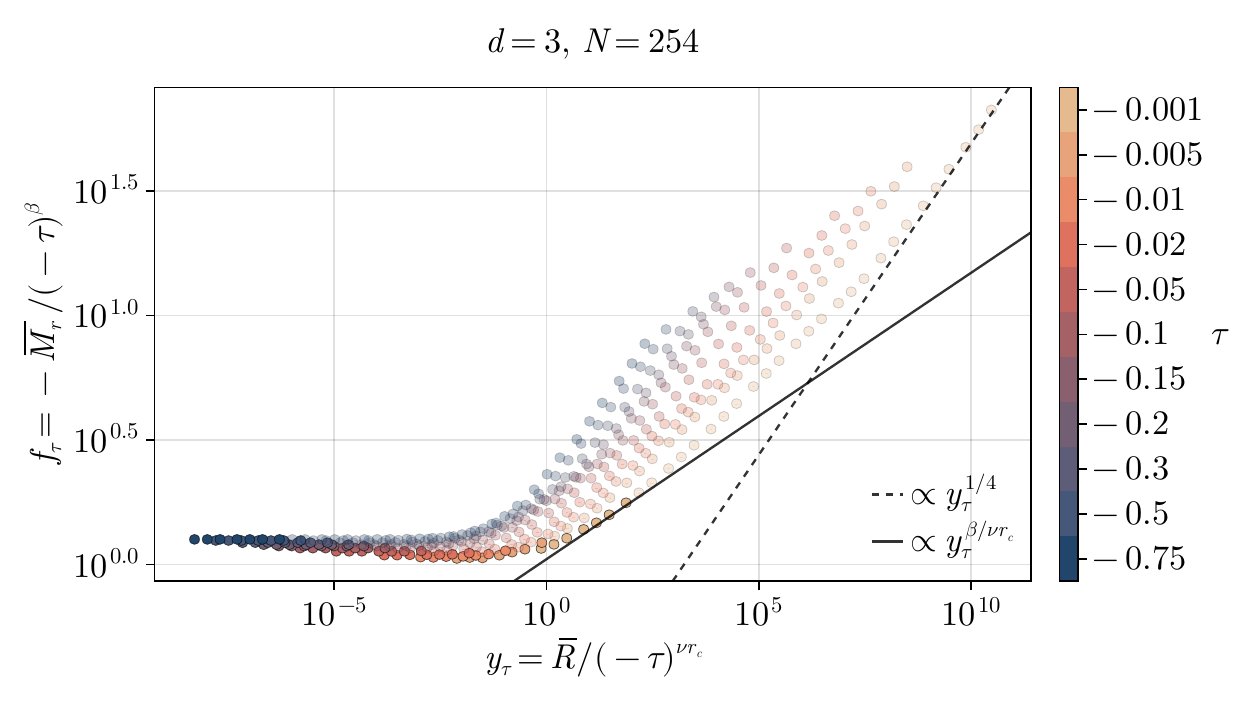}
    \includegraphics[width=.498\textwidth]{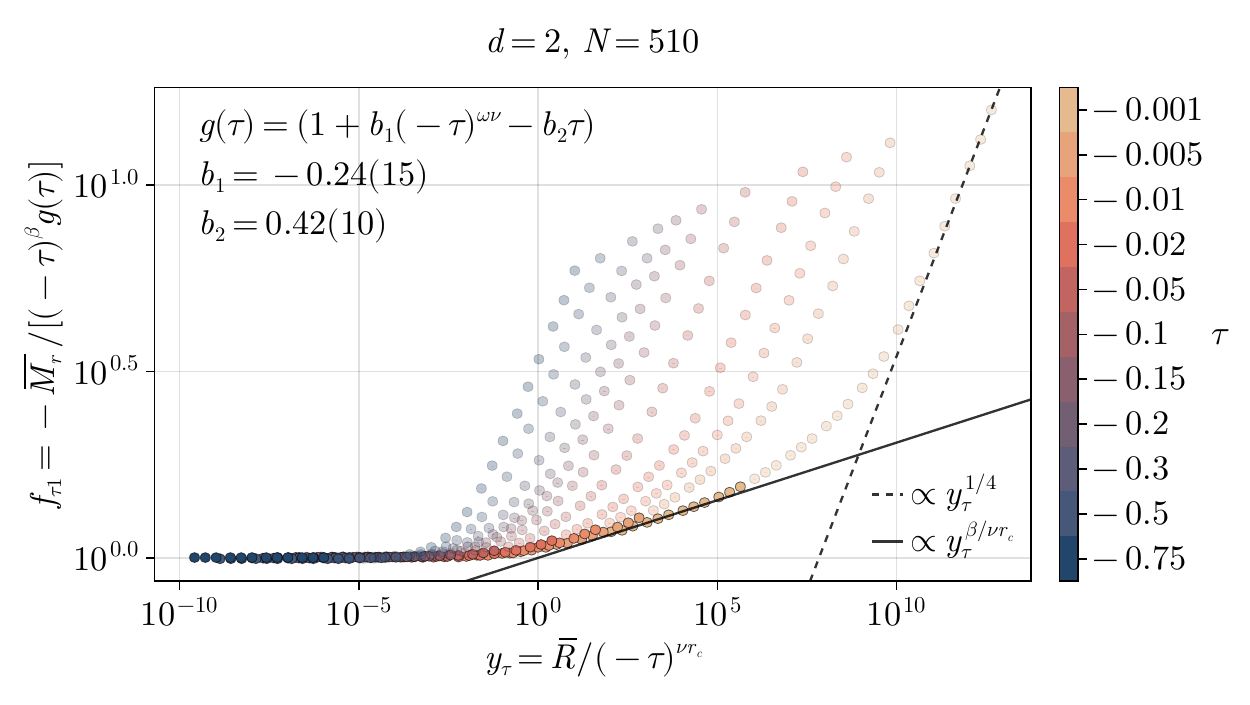}
    \includegraphics[width=.498\textwidth]{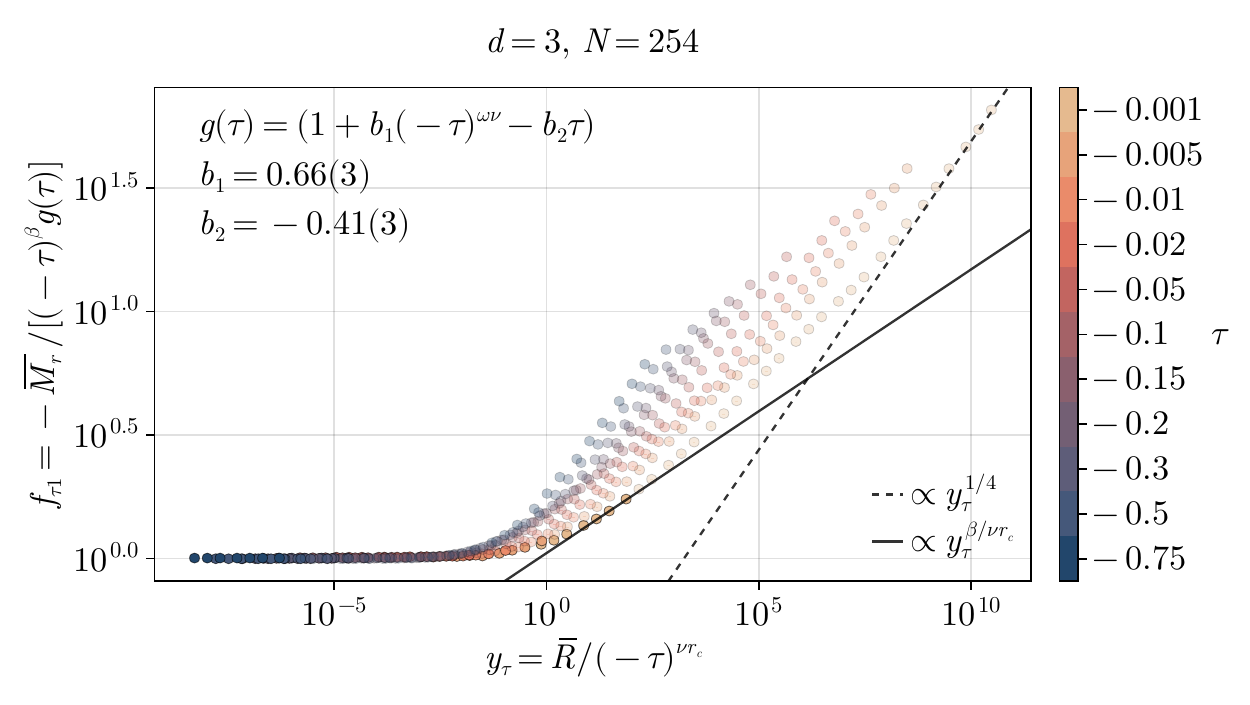}
    \caption{\label{fig:scaled_Mr_vs_R} Scaled remanent magnetization in $d=2$ (left) and $d=3$ (right) spatial dimensions.
        Data points corresponding to fast quenches ($R>10^{-3}$ for $d=2$ and $R>2.5\times 10^{-5}$ for $d=3$) are drawn semi-transparent.
        The solid black lines show the critical  finite-time scaling behavior, while the black dashed lines indicate the asymptotic mean-field scaling behavior of the underdamped system.
    }
\end{figure}

According to the scaling ansatz in \cref{eq:Mr_scaling} the universal scaling functions $f_\tau(y_\tau)$ and $f_R(x_R)$ can be obtained via appropriate rescaling of the magnetization and the external control parameters.
We present the rescaled remanent magnetization leading to the former scaling function in the upper panels of \cref{fig:scaled_Mr_vs_R}.
Although we find a finite-time scaling window of quench rates where the data matches the expected critical scaling behavior, indicated by the solid black line, this scaling window is relatively small.
Especially in $d=3$ dimensions, the scaled magnetization data for all temperatures except the one closest to the critical point, do not appear to follow the critical scaling behavior, but instead immediately cross over from critical equilibrium scaling to mean-field scaling as the quench rate is increased.
We were not able to significantly improve the scaling collapse by including the leading finite-time scaling correction~$\sim R^{\omega/r_c}$ as was done for the coercive field which we therefore omit in \cref{fig:scaled_Mr_vs_R}.

However, the equilibrium scaling collapse in the slow quench region can be significantly improved by including the leading equilibrium scaling correction. As previous studies have shown, the spontaneous magnetization is well described by the following ansatz including corrections to scaling~\cite{Engels:2003,Schweitzer:2020noq}
\begin{equation}
    \bar M_{eq}(\tau) = (-\tau)^{\beta} \left[1 + b_1 (-\tau)^{\omega \nu} + b_2 (-\tau)\right],
    \label{eq:equilibrium_scaling_correction}
\end{equation}
with non-universal constants $b_1$ and $b_2$ that are determined from a single fit to the numerical data at the slowest available quench rate.
Taking the equilibrium scaling correction into account, the remanent magnetization is expected to behave as
\begin{equation}
    \bar M_r(\tau, R) \approx (-\tau)^{\beta} \left[1 + b_1 (-\tau)^{\omega \nu} + b_2 (-\tau)\right] f_{\tau 1}\left(\bar R / (-\tau)^{\nu r_c}\right),
    \label{eq:Mr_scaling_with_equilibrium_correction}
\end{equation}
which we find to result in a perfect collapse of the data for slow quenches as shown in the lower panels of \cref{fig:scaled_Mr_vs_R}.

\subsection{Scaling functions for the magnetization}
Having established the scaling behavior of two characteristic points of the magnetization during the transition, we will now present results for the general behavior of the full two-parameter finite-time scaling function $f_R(x_R, y_R)$ as defined in \cref{eq:fts}.
This function fully describes the universal out-of-equilibrium scaling behavior of the magnetization in quenches across the first-order transition line in the vicinity of the critical point, and can be obtained by rescaling the numerical magnetization data in appropriate ranges of temperatures and quench rates.
Based on the observation made in the previous sections, that corrections to finite-time scaling become relevant in fast quenches above a threshold of roughly $R \gtrsim 10^{-3}$ for $d=2$ and $R \gtrsim 2.5\times 10^{-5}$ for $d=3$, we will limit the data we consider in the following analysis to sufficiently slow quenches below this identified model-dependent threshold.

\begin{figure}[tb]
    \centering
    \includegraphics[width=.8\textwidth]{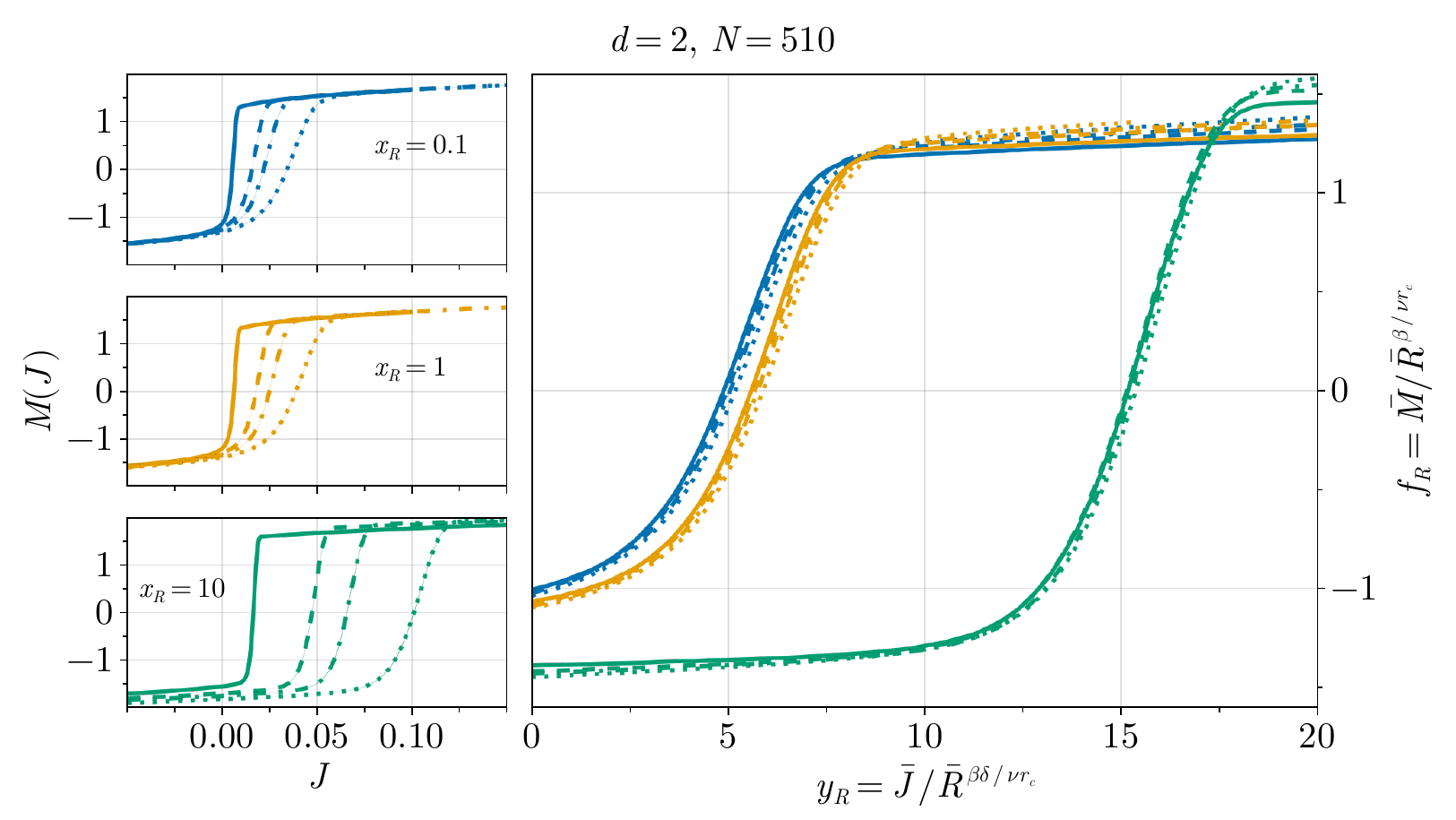}
    \caption{\label{fig:M_vs_J_constant_xR}
    Left: Magnetization $M$ as a function of the external field $J$ for different quench rate and temperature combinations corresponding to fixed values of the scaling variable $x_R = -\tau / \bar R^{1/\nu r_c}$; $x_R=0.1$ (blue), $x_R=1$ (yellow) and $x_R=10$ (green).
    Different quench rates are represented by different line styles:
    \linesym{solid} $R=10^{-5}$,\; \linesym{dashed} $R=10^{-4}$,\; \linesym{dash dot} $R=2\times10^{-4}$,\; \linesym{dotted} $R=5\times10^{-4}$.
    The reduced temperature $\tau$ was chosen accordingly to arrive at the constant values of $x_R$ presented above.
    Right: Rescaled magnetization $f_R = \bar M / \bar R^{\beta/\nu r_c}$ as a function of the scaled external field $y_R = \bar J / \bar R^{\beta\delta/\nu r_c}$.
    Different curves of identical color are expected to collapse onto common curves representing slices at fixed $x_R$ of the universal two-parameter scaling function $f_R(x_R,y_R)$ defined in \cref{eq:fts}.
    }
\end{figure}

For quenches across the critical point, corresponding to the limit $x_R \to 0$, the scaling function $f_R(x_R=0,y_R)$  in two and three spatial dimensions was already obtained in~\cite{Sieke:2024dns}.
To see how this function changes with increasing values of the scaling variable $x_R = -\tau / \bar R^{1/\nu r_c}$, we first investigate slices at larger fixed values of $x_R \in \{0.1, 1, 10\}$.
For this purpose, we performed simulations of transitions in $d=2$ for different quench rates $R$ in the identified dynamic scaling regime ($R \in \{10^{-5}, 10^{-4}, 2 \times 10^{-4}, 5 \times 10^{-4} \}$) and selected the reduced temperature $\tau$ accordingly to obtain the desired constant values of $x_R$.
The behavior of the magnetization in these transitions is shown in \cref{fig:M_vs_J_constant_xR}.

On the left-hand side of \cref{fig:M_vs_J_constant_xR} we present numerical results for the magnetization as a function of the external driving field $J(t)$ for four different quench rates $R=10^{-5}$ (solid lines), $R=10^{-4}$ (dashed lines), $R=2 \times 10^{-4}$ (dash-dotted lines) and $R= 5 \times 10^{-4}$ (dotted lines), with appropriately chosen reduced temperatures $\tau$ such that the scaling variable $x_R$ takes on a value of $x_R = 0.1$ (blue curves), $x_R = 1$ (yellow curves) and $x_R = 10$ (green curves) respectively.

The right-hand side of \cref{fig:M_vs_J_constant_xR} shows the rescaled magnetization as a function of the scaling variable $y_R = \bar J / \bar R^{\beta\delta/\nu r_c}$ according to the finite-time scaling ansatz of \cref{eq:fts}.
We find that the different quench rate and temperature combinations corresponding to identical fixed values of $x_R$ indeed collapse together as expected.
Although the data collapse is not perfect, with slight deviations being visible for the largest value of $R = 5 \times 10^{-4}$ (dotted lines), the scaling collapse is generally good throughout the whole range of the magnetization curves, and not only limited to the special points $M=0$ and $J=0$ defining the coercive field and remanent magnetization respectively, which were investigated in the previous sections.

The curves corresponding to values of $x_R = 0.1$ (blue), $x_R = 1$ (yellow) are observed to be rather similar to each other, while the curves corresponding to $x_R = 10$ (green) differ significantly from the other two sets of curves.
We therefore expect the magnetization in sufficiently fast first-order transitions, such that $x_R \lesssim 1$, to also behave very similar to quenches directly across the critical point, as the scaling function $f_R(x_R,y_R)$ does not seem to exhibit a strong dependence on the scaling variable $x_R$ in this regime.

However, the fact that the finite-time scaling regime itself is limited to slow quenches due to the inevitable crossover to mean-field behavior in rapid quenches, gives rise to a lower bound of the reduced temperature $\tau$ for which close-to-critical finite-time scaling behavior can be observed.
Considering the observed upper limits for the quench rate of $R \lesssim 10^{-3}$ in $d=2$, and $R \lesssim 2.5 \times 10^{-5}$ in $d=3$ of our model, the corresponding lower bounds for the reduced temperature $\tau$ for which $x_R= -\tau / \bar R^{1/\nu r_c} \lesssim 1$ can be estimated to be $\tau \gtrsim -0.025$ for $d=2$ and $\tau \gtrsim -0.005$ for $d=3$.
Importantly, these bounds depend entirely on non-universal amplitudes and the microscopic details of the model under consideration, and can thus not be transferred to other systems in the same universality class.

\begin{figure}[tb]
    \centering
    \includegraphics[width=.48\textwidth]{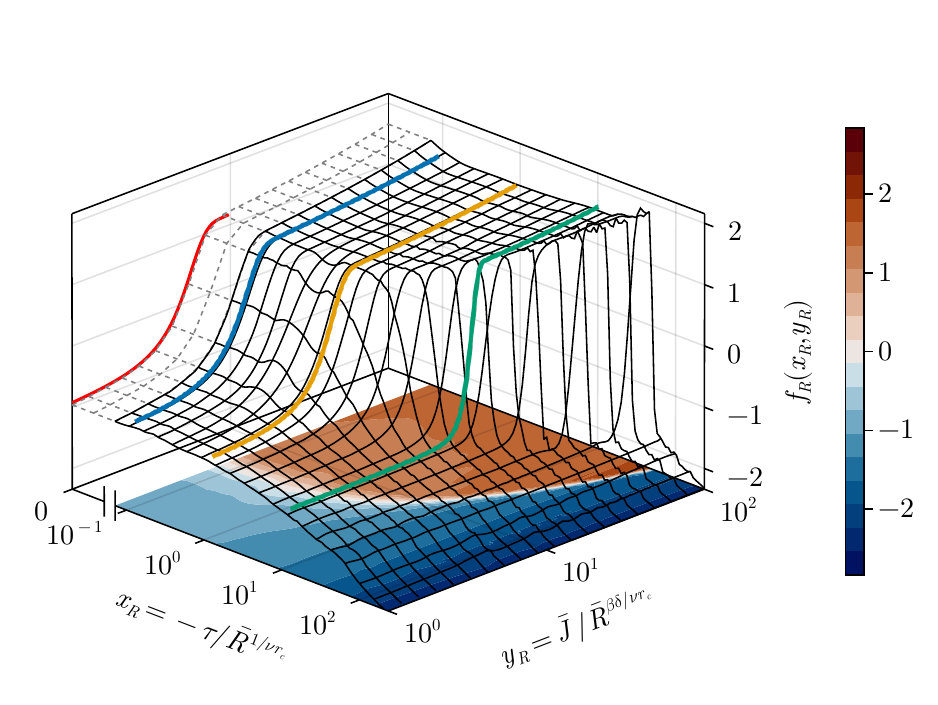}
    \includegraphics[width=.48\textwidth]{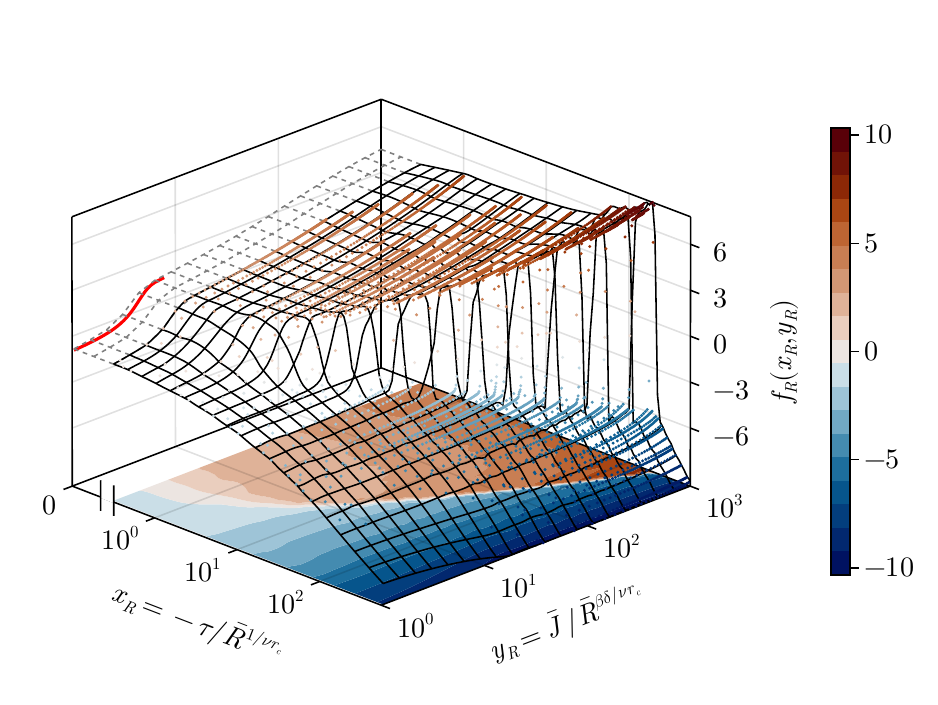}
    \caption{
    \label{fig:interpolated_grid_256_with_points}
    Two-parameter scaling collapse of the magnetization in $d=2$ (left) and $d=3$ (right) spatial dimensions.
    The solid wireframe represents a polyharmonic spline interpolation of the rescaled data in the finite-time scaling regime, revealing the universal scaling function $f_R(x_R,y_R)$ defined in \cref{eq:fts}.
    The dashed wireframe shows a constant extrapolation of the scaling function at the smallest available value of $x_R$ to the limit $x_R \to 0$, which is compared to the critical scaling function obtained in~\cite{Sieke:2024dns} shown as a solid red curve.
    In the left panel for the two-dimensional case, the colored curves correspond to the $R=10^{-5}$ data at fixed values of $x_R=0.1$ (blue) $x_R=1.0$ (yellow) and $x_R=10.0$ (green) that were presented in \cref{fig:M_vs_J_constant_xR}.
    In the right panel for the three-dimensional case, we instead show the data points of the magnetization that were used in the interpolation.
    }
\end{figure}

To qualitatively visualize the behavior of the scaling function $f_R(x_R,y_R)$ in the continuous range of the two scaling variables, we rescaled all of our magnetization data in the finite-time scaling regime according to the ansatz in \cref{eq:fts}, and performed a two-dimensional polyharmonic spline interpolation revealing the underlying universal scaling function, which is shown as a solid wireframe in the left and right panel of \cref{fig:interpolated_grid_256_with_points} for the $d=2$ and $d=3$ cases respectively.
The solid colored lines in the left panel of \cref{fig:interpolated_grid_256_with_points} correspond to the $R=10^{-5}$ data that was presented in \cref{fig:M_vs_J_constant_xR}.
We additionally performed a constant extrapolation of the numerically obtained scaling function $f_R$ to the $x_R \to 0$ limit, visualized by a dashed wireframe, to compare with the critical scaling functions that were previously obtained in~\cite{Sieke:2024dns} which are shown as solid red lines in \cref{fig:interpolated_grid_256_with_points}.
We find these critical scaling functions to match the two-parameter scaling function $f_R(x_R,y_R)$ for small values of $x_R$, supporting the idea that in sufficiently fast quenches across the first-order transition, the magnetization can behave very similar to quenches directly across the critical point.

The transition between the two ordered phases changes from a very gradual transition in fast quenches close to the critical point, corresponding to small values of $x_R$, to a much sharper transition in slower quenches further away from the critical point, corresponding to larger values of $x_R$.
This behavior is more pronounced in the $d=3$ case shown in the right panel of \cref{fig:interpolated_grid_256_with_points}, due to the larger value of the critical exponent $\beta$ associated with the spontaneous magnetization which governs the asymptotic behavior of $f_R(x_R, y_R)$ for large $x_R$.
There we also show the data points of the scaled magnetization that were used to construct the interpolation function.
Although the out-of-equilibrium magnetization in the finite-time scaling regime still strongly depends on all three control parameters, the reduced temperature $\tau$, the time-dependent symmetry-breaking field $J$, and the quench rate $R$, it is found to be very well described by a universal function $f_R(x_R, y_R)$ of only two scaling  variables.

\begin{figure}[tb]
    \includegraphics[width=.498\textwidth]{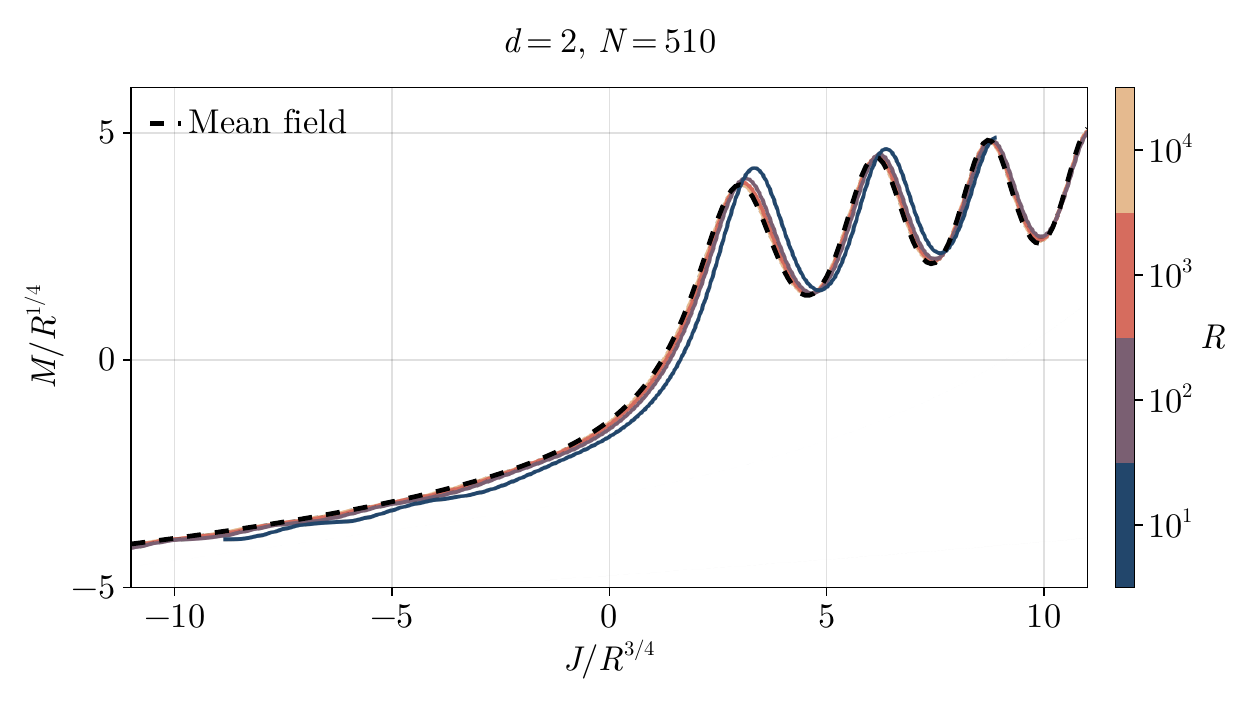}
    \includegraphics[width=.498\textwidth]{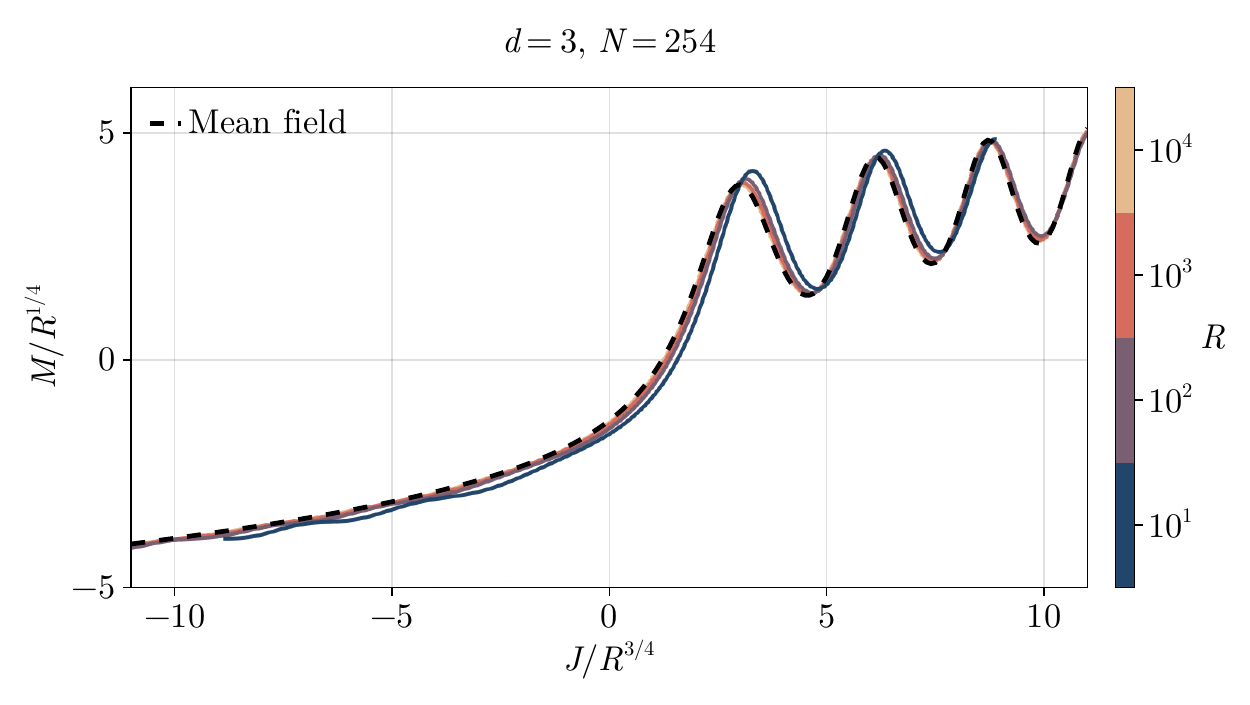}
    \caption{\label{fig:M_vs_J_MF_limit}
        Scaling collapse of the magnetization to the underdamped mean-field scaling function (dashed black line) for fast quenches ($R \in \{10^{1}, 10^{2}, 10^{3}, 10^{4} \}$).
        Colored lines show the numerically obtained magnetization curves in quenches close to the critical point ($\tau=-0.001$) including fluctuations, scaled with the mean-field exponents identified in \cref{sec:mean_field_results}.
        In both, two and three dimensions (left and right panels respectively), the scaled magnetization collapses perfectly to the mean-field scaling function for quench rates $R \gtrsim 10^2$.
    }
\end{figure}

In a complementary regime of very rapid quenches, we find temperature independent mean-field scaling behavior, as was already observed for the coercive field and remanent magnetization in \cref{fig:Jc_vs_R,fig:Mr_vs_R} respectively.
In this regime, the magnetization is well described by a mean-field scaling function that is completely independent of temperature and dimensionality as shown.
The scaling collapse to this function, is shown in \cref{fig:M_vs_J_MF_limit} where we present the scaled magnetization data that was obtained fully including fluctuations in near-critical quenches at $\tau = -0.001$ where deviations from mean-field behavior are expected to be the largest.
For quench rates $R \gtrsim 10^{2}$, the magnetization data, scaled with mean-field exponents for underdamped Langevin dynamics as identified in \cref{sec:mean_field_results}, is found to be in very good agreement with the mean-field scaling function represented by the dashed black curves.

\section{Conclusions}
\label{sec:conclusions}

We have studied the out-of-equilibrium behavior of a relativistic scalar field theory with a $Z_2$ order-parameter symmetry and dissipative dynamics in first-order phase transitions driven by a linearly varying symmetry-breaking field.
By investigating a wide range of temperatures and quench rates, we identified two distinct dynamic regimes where scaling behavior emerges and computed the corresponding universal scaling functions for the order-parameter in the respective regimes.

In sufficiently fast quenches, mean-field finite-time scaling behavior of the magnetization was observed, even close to the critical temperature, which was verified by comparing our numerical results to the expected theoretical mean-field exponents as well as results from direct numerical mean-field simulations.
The mean-field exponents differ from the usual ones for the overdamped Langevin equation, which is commonly used to describe the critical dynamics of Ising systems, because of the inertial term in our second order Langevin equations of motion dominating in the fast quench regime.
Asymptotic slow quenches result in non-universal behavior dominated by nucleation and growth.
\ref{sec:spinodal_like_scaling} provides a brief discussion of our results with respect to the recently proposed spinodal-like scaling in this regime~\cite{Pelissetto:2025pzi}.
If however, the quench rate is sufficiently large to avoid classical tunneling via droplet nucleation but still small enough for spatial correlations to form before the transition, Kibble-Zurek like scaling behavior consistent with the expected critical exponents for the $d=2$ and $d=3$ Ising universality classes with Model A dynamics can emerge even in field-driven first-order phase transitions.
Deviations from critical non-equilibrium scaling, due to a finite reduced temperature $\tau$, are observed to be well described in terms of the dimensionless scaling variable $x_R = -\tau / \bar R^{1/\nu r_c}$.
At the transition point where the magnetization changes sign characterized by the coercive field $J_c$, the scaling collapse was found to be significantly improved by including the leading algebraic correction in accordance with the established finite-time scaling theory.

The dynamics studied in this work may be broadly relevant to a wide range of physical settings, such as first-order cosmological phase transitions which are essential in electroweak baryogenesis and provide a potential source of gravitational waves in the early universe~\cite{Moore:2000jw,Caprini:2019egz,Gould:2022ran,Sagunski:2023ynd,vandeVis:2025efm}. Near critical and first-order quenches can also be relevant for the QCD phase diagram in the high net-baryon density region as probed in heavy-ion collision experiments at lower beam energies~\cite{Bzdak:2019pkr}.
A more direct connection requires extending the present framework to incorporate the specific dynamics of these systems, taking into account the relevant conservation laws and reversible mode couplings \cite{Roth:2024rbi,Roth:2024hcu}.
Recent progress in this direction includes e.g.\ numerical simulations of the out-of-equilibrium dynamics of Model G, the dynamic universality class relevant for the $O(4)$ chiral transition~\cite{Florio:2025zqv,Florio:2025lvu}, and simulations of stochastic fluid dynamics in the Model H universality class governing the non-equilibrium dynamics near the conjectured QCD critical point~\cite{Chattopadhyay:2024jlh,Chattopadhyay:2024bcv}. As a natural next step in our present classical-statistical simulation framework, it will be particularly interesting to consider energy conservation in order to study the effects of the latent heat on the non-equilibrium dynamics in the first-order quenches.

\section*{CRediT Authorship Contribution Statement}
\textbf{Leon Sieke:} Conceptualization, Data Curation, Formal Analysis, Investigation, Methodology, Software, Visualization, Writing - Original Draft.
\textbf{Jessica Fuchs:} Investigation, Software, Validation, Visualization, Writing - Reviewing \& Editing.
\textbf{Lorenz von Smekal:} Conceptualization, Funding Acquisition, Methodology, Project Administration, Resources, Supervision, Writing - Reviewing \& Editing.

\section*{Declaration of competing interest}
The authors declare that they have no known competing financial interests or personal relationships that could have appeared to influence the work reported in this paper.

\section*{Data availability}
Data will be made available upon request.

\section*{Acknowledgments}
We thank Sören Schlichting, Johannes Roth, Guy Moore and Adrien Florio for insightful discussions.
We acknowledge Kari Rummukainen and Fan Zhong for drawing our attention to related literature.
This work was supported by the Deutsche Forschungsgemeinschaft (DFG, German Research Foundation) through the CRC-TR 211 `Strong-interaction matter under extreme conditions' -- project number 315477589 -- TRR 211.

\appendix

\section{Additional scaling functions for the coercive field and remanent magnetization}
\label{sec:additional_scaling_functions}

\begin{figure}[htb]
    \includegraphics[width=.498\textwidth]{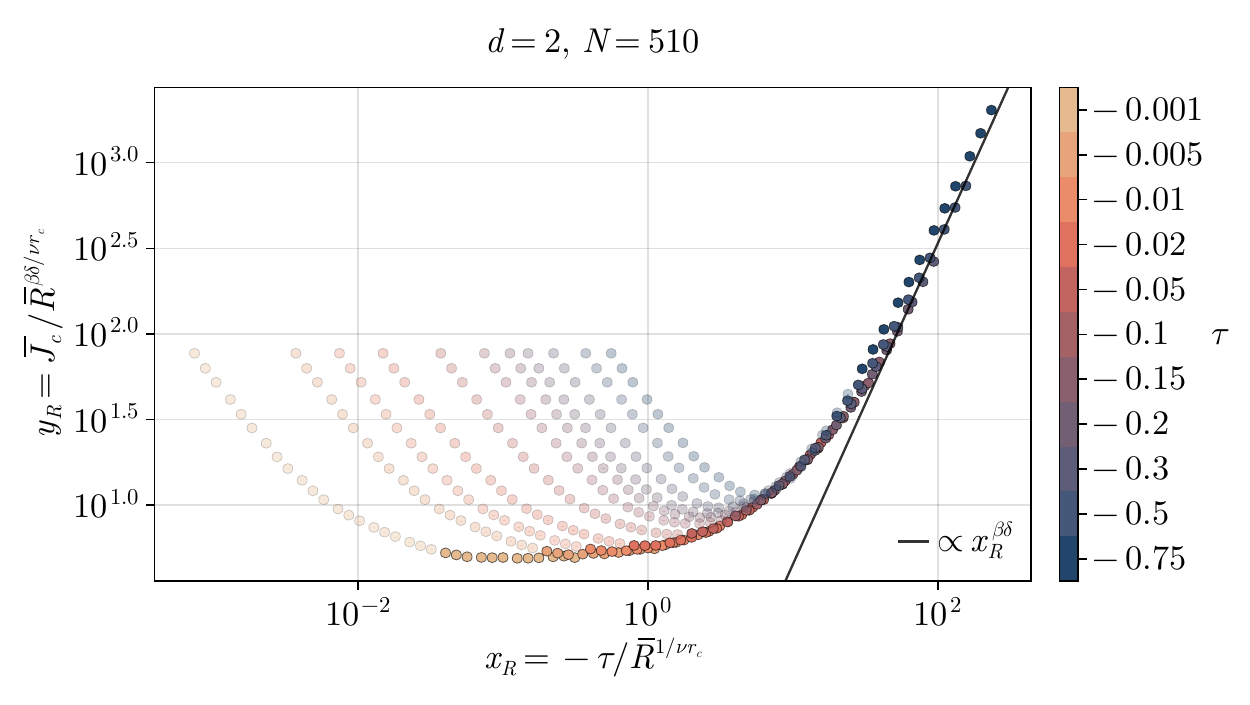}
    \includegraphics[width=.498\textwidth]{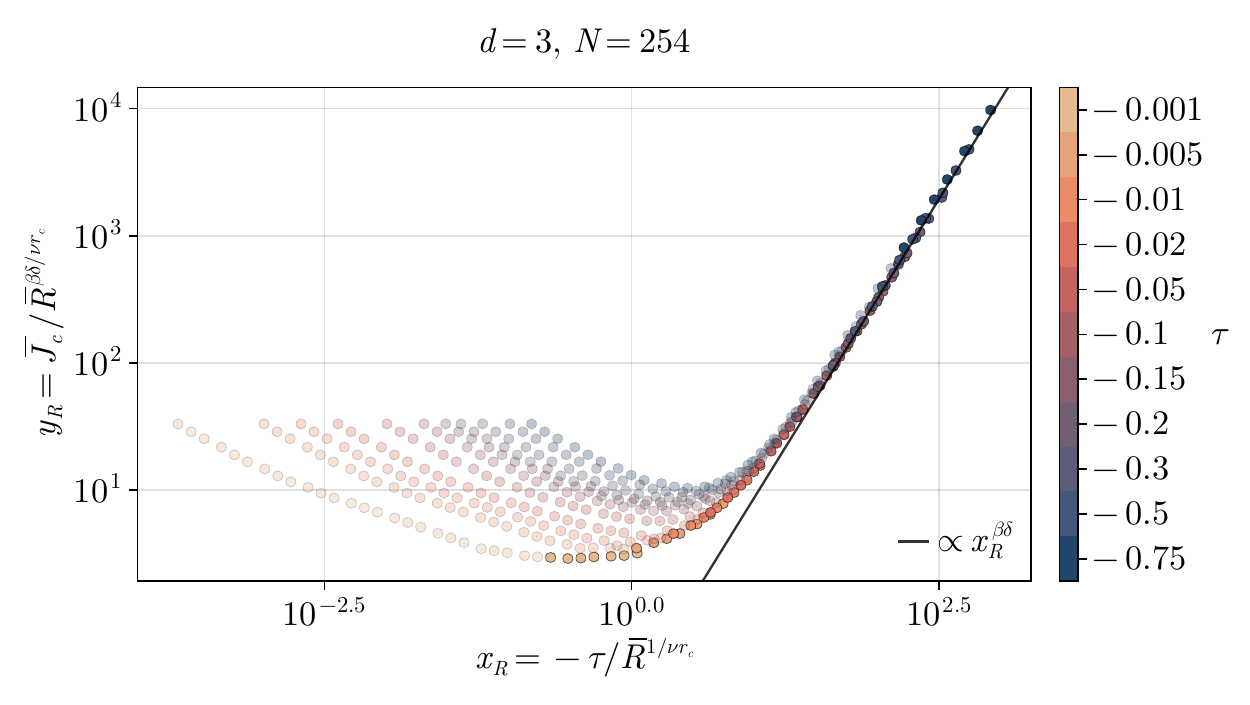}
    \includegraphics[width=.498\textwidth]{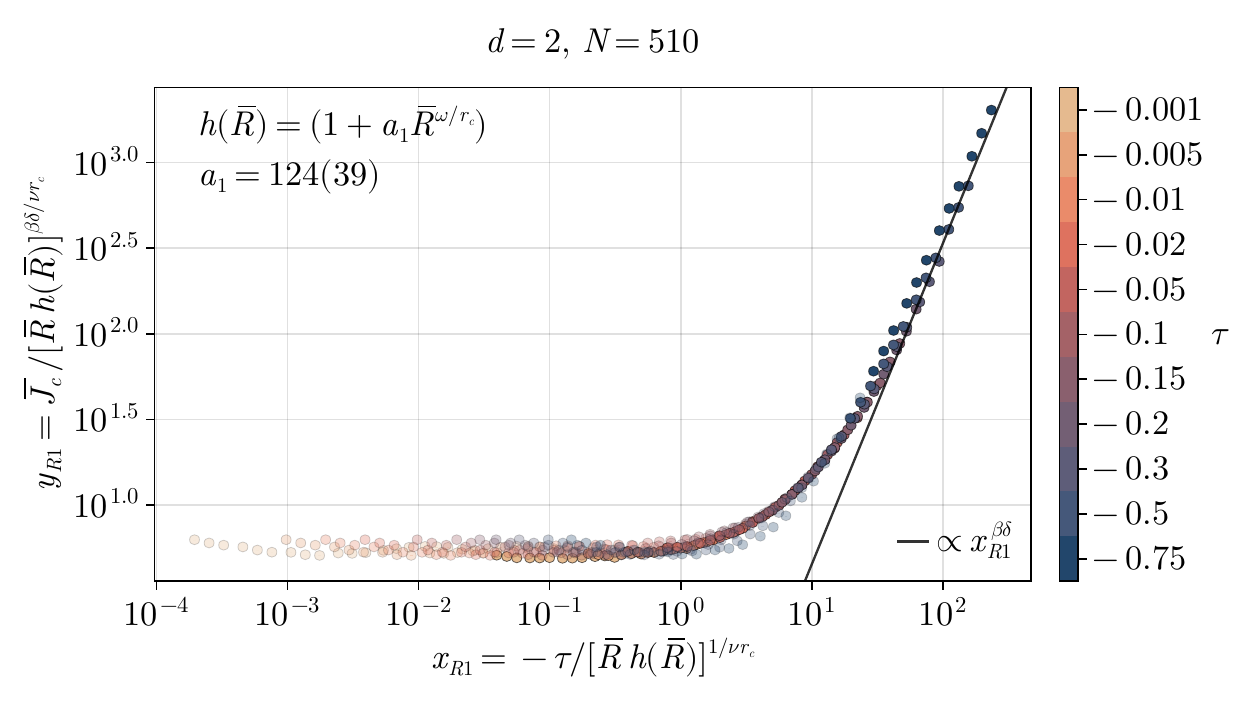}
    \includegraphics[width=.498\textwidth]{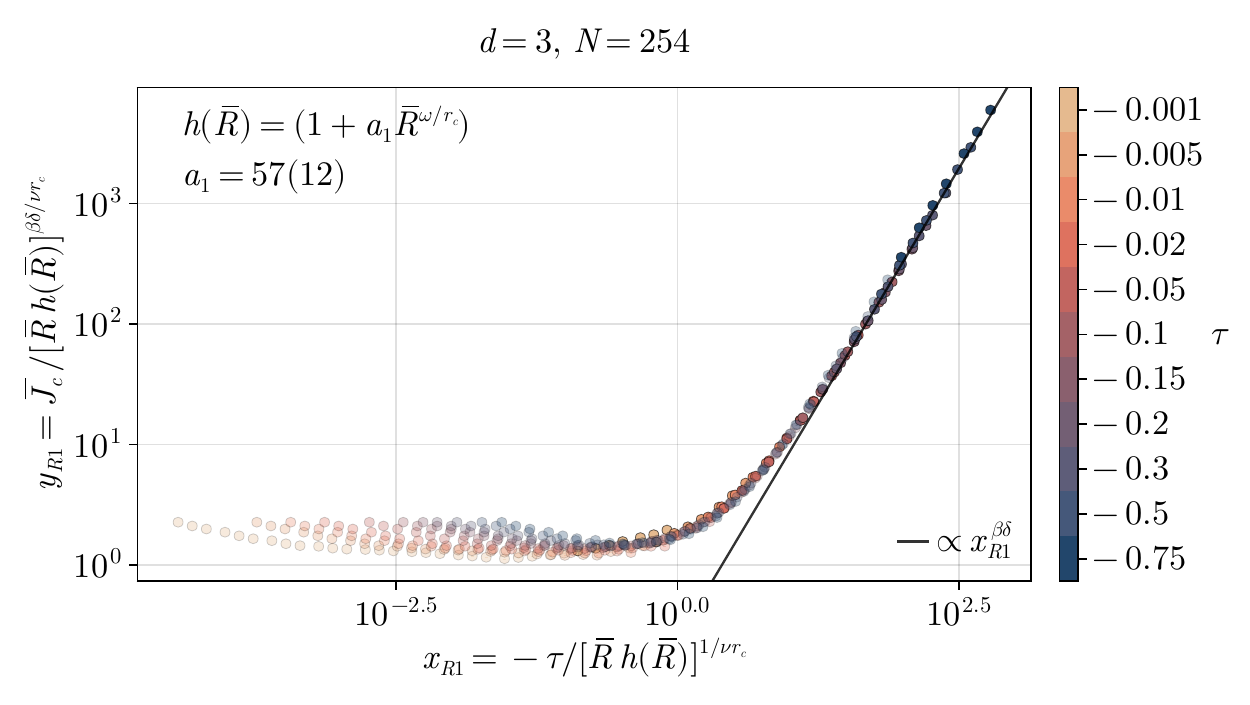}

    \caption{\label{fig:scaled_Jc_vs_R_fts} Scaling collapse of the coercive field in finite-time scaling form for $d=2$ (left) and $d=3$ (right) spatial dimensions. Data points corresponding to fast quenches ($R>10^{-3}$ for $d=2$ and $R>2.5\times 10^{-5}$ for $d=3$) are drawn semi-transparent. The solid black line indicates the expected asymptotic behavior in the equilibrium limit $R \to 0$. The upper panels show the rescaled data according to \cref{eq:Jc_fts_scaling_form}, while the lower panels show the rescaled data including the leading finite-time scaling correction according to \cref{eq:Jc_fts_scaling_form_with_correction}.
    }
\end{figure}

In this section we present additional scaling functions in an alternative form to the ones presented in \cref{fig:scaled_Jc_vs_R,fig:scaled_Mr_vs_R}, following the definition in \cref{eq:fts}.
While these contain the same information and can be transformed into each other via the relationship shown in \cref{eq:fR_from_fTau}, they allow to visualize the scaling behavior in a different way, highlighting separate dynamic regimes.
\cref{fig:scaled_Jc_vs_R_fts} shows the finite-time scaling function for the coercive field $J_c$.
In this form, the cross-over from critical to mean-field scaling behavior is resolved in more detail.
In the lower panels of \cref{fig:scaled_Jc_vs_R_fts} we again include the leading finite-time scaling correction, which is of the form
\begin{equation}
    \bar J_c(\tau, R) \approx [\bar R(1 + a_1 \bar R^{\omega/r_c})]^{\beta\delta/\nu r_c} y_{R1}(-\tau / [\bar R([1 + a_1 \bar R^{\omega/r_c}])]^{1/\nu r_c} ).
    \label{eq:Jc_fts_scaling_form_with_correction}
\end{equation}
The best-fit values for the non-universal constant $a_1$ are identical to the ones shown previously in \cref{fig:scaled_Jc_vs_R}.

\begin{figure}[htb]
    \includegraphics[width=.498\textwidth]{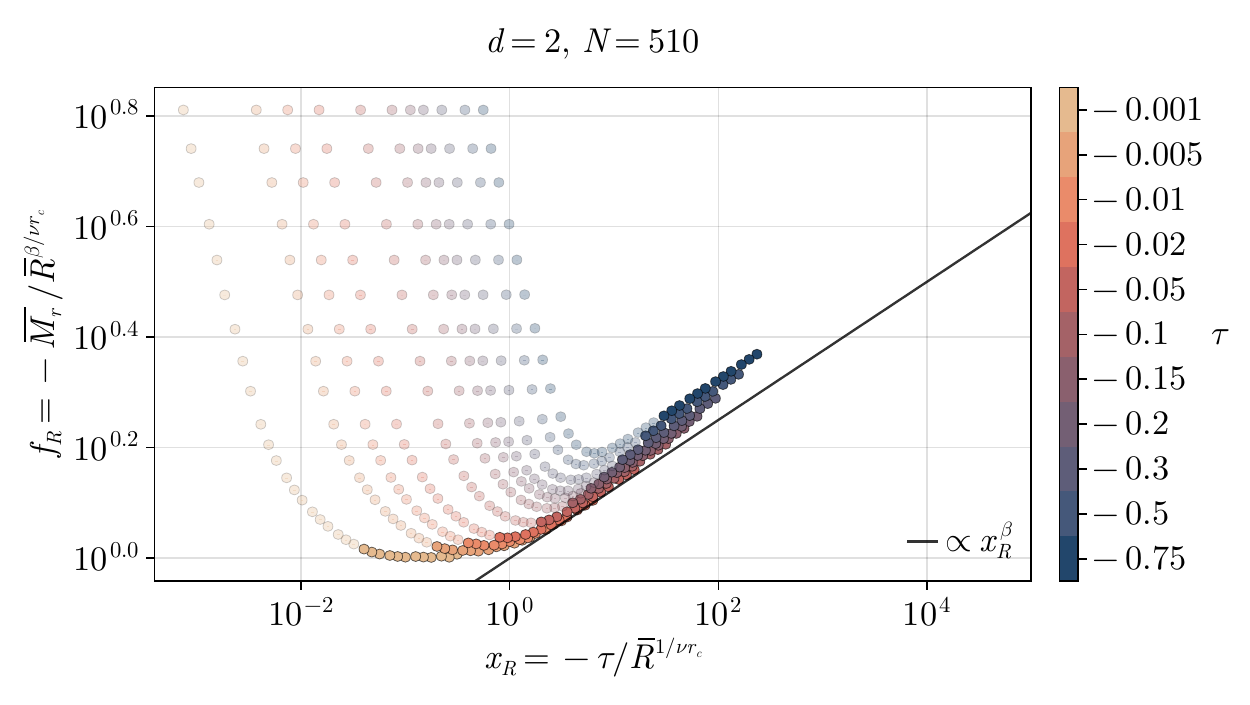}
    \includegraphics[width=.498\textwidth]{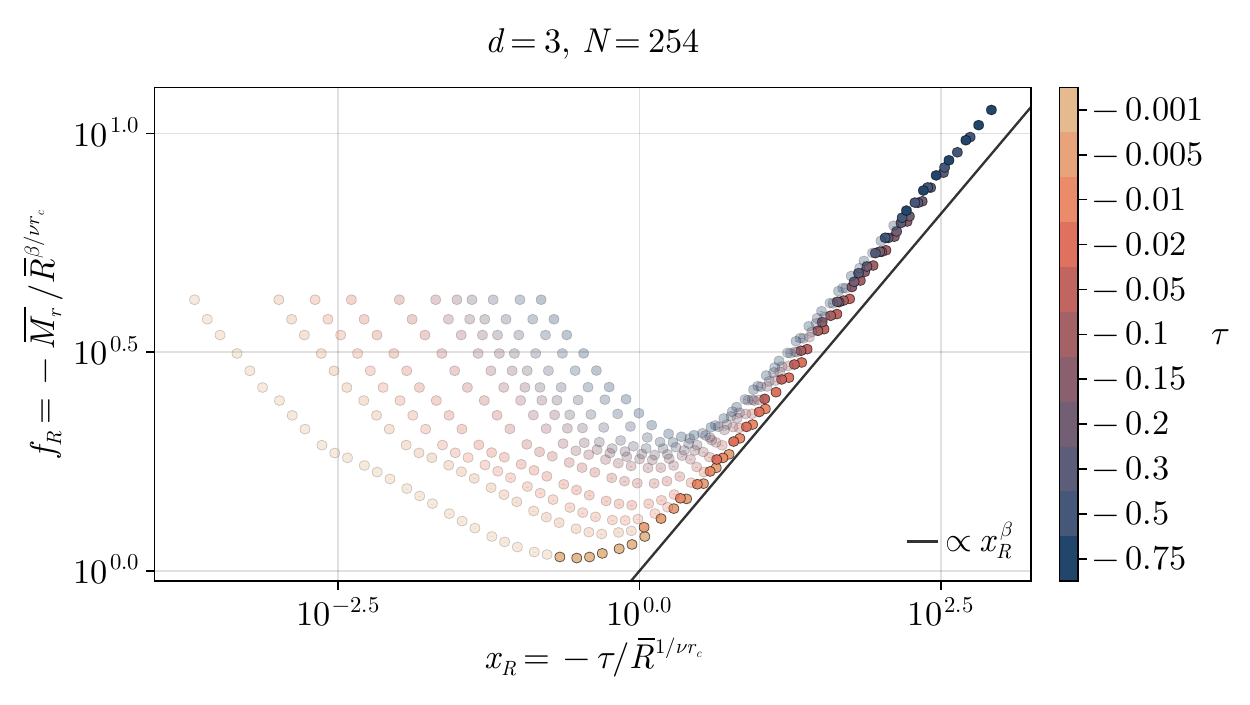}
    \includegraphics[width=.498\textwidth]{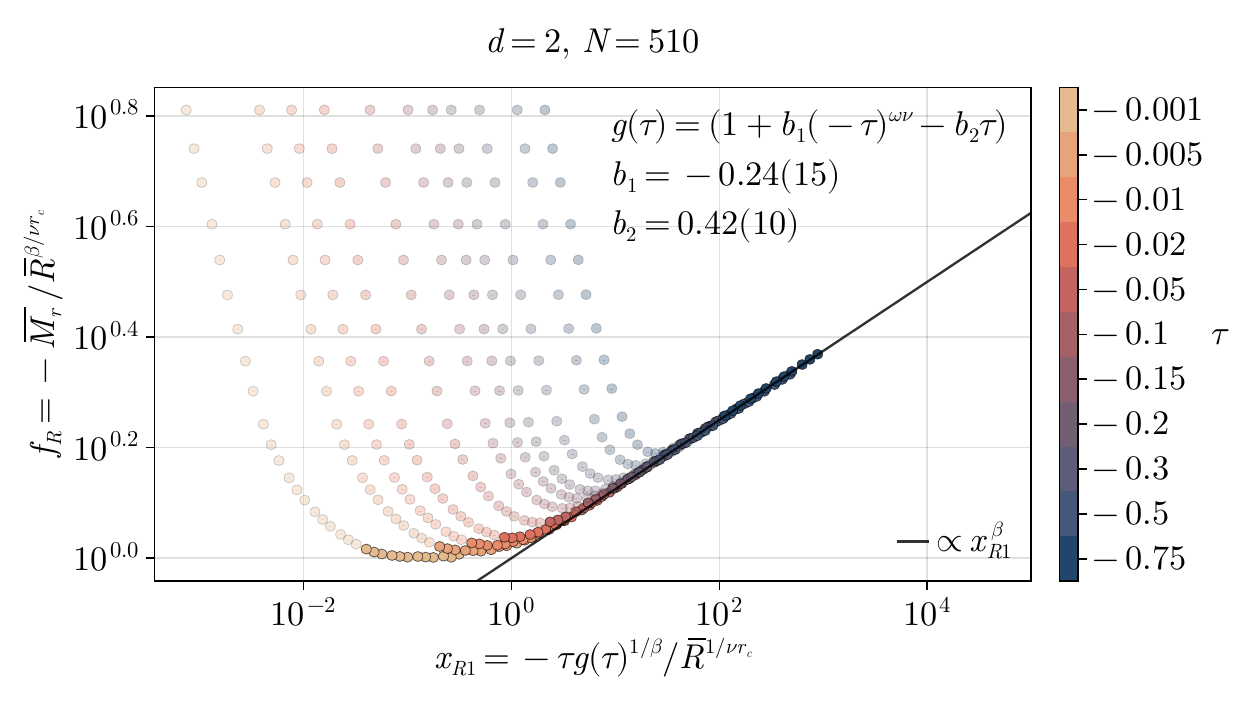}
    \includegraphics[width=.498\textwidth]{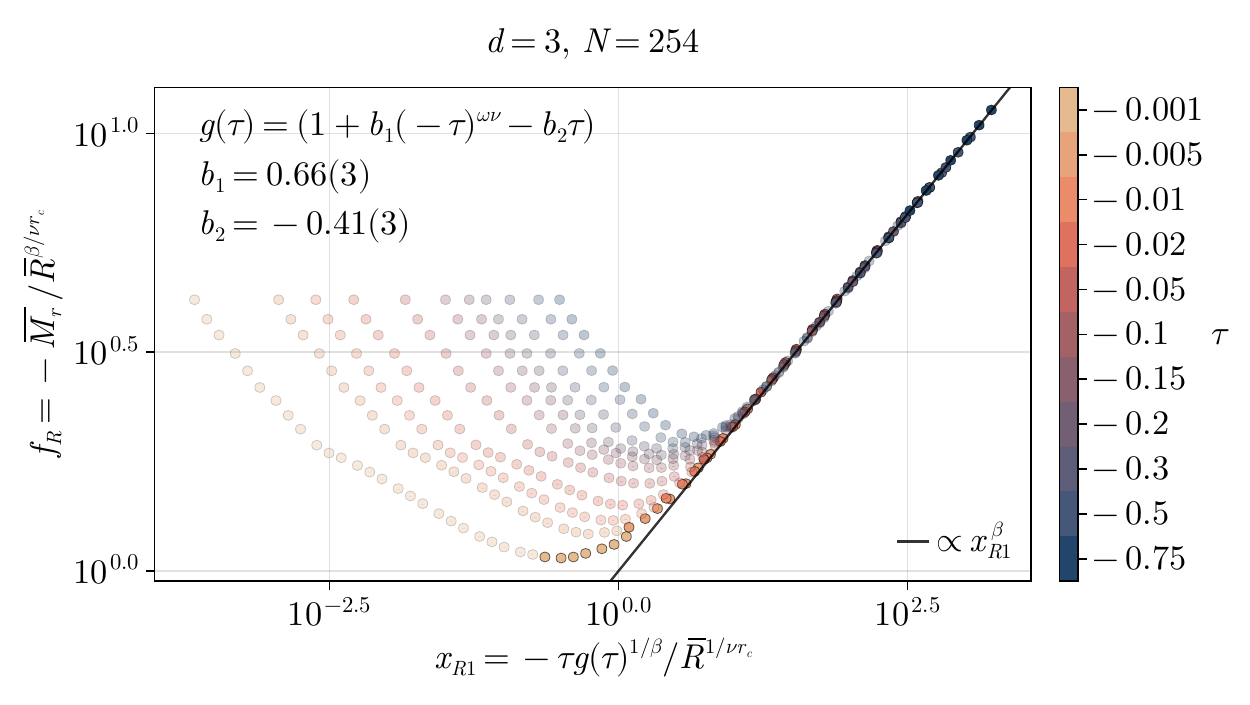}
    \caption{\label{fig:scaled_Mr_vs_R_fts} Scaling collapse of the remanent magnetization in finite-time scaling form for $d=2$ (left) and $d=3$ (right) spatial dimensions.
        Data points corresponding to fast quenches ($R>10^{-3}$ for $d=2$ and $R>2.5\times 10^{-5}$ for $d=3$) are drawn semi-transparent.
        The solid black lines indicate the asymptotic behavior in the equilibrium limit $R \to 0$.
        The upper panels show the rescaled data according to \cref{eq:Mr_scaling}, while the lower panels show the rescaled data including the leading equilibrium scaling correction according to \cref{eq:Mr_fts_scaling_with_equilibrium_correction}
    }
\end{figure}

Correspondingly, the rescaled remanent magnetization in finite-time scaling form is given by
\begin{equation}
    \bar M_r(\tau, R) \approx \bar R^{\beta/\nu r_c} f_{R 1}\left(- \tau \left[1 + b_1 (-\tau)^{\omega \nu} + b_2 (-\tau)\right]^{1/\beta}  / \bar R^{1/\nu r_c}\right),
    \label{eq:Mr_fts_scaling_with_equilibrium_correction}
\end{equation}
which immediately follows from \cref{eq:Mr_scaling_with_equilibrium_correction} due to the relationship between the scaling functions in \cref{eq:fR_from_fTau}.
The rescaled remanent magnetization in this form, with and without equilibrium scaling correction is shown in the lower and upper panels of \cref{fig:scaled_Mr_vs_R_fts} respectively.

\section{On spinodal-like scaling behavior in slow quenches far from the critical point}
\label{sec:spinodal_like_scaling}

In this section, we address the regime of very slow quenches far away from the critical point.
The phase transitions in this regime are neither described by the two-parameter finite-time scaling function presented in \cref{fig:interpolated_grid_256_with_points}, nor by the mean-field scaling function shown in \cref{fig:M_vs_J_MF_limit}.
Instead, a recent study of the out-of-equilibrium behavior of Ising systems in slowly driven magnetic first-order transitions proposed scaling behavior in terms of the scaling variable
\begin{equation}
    \sigma = J(t) (\ln t)^{\kappa}, \quad\text{with } \kappa = d/(d-1).
    \label{eq:multi_droplet_scaling}
\end{equation}
This scaling variable was argued to arise based on a previously identified finite-size-scaling variable  $\Phi = J(t) L^d$ and the two assumptions that the relevant length scale for the nucleation-initiated transition is the time-dependent typical droplet size $\mathcal{R}(t)$, and the time to create a droplet scales exponentially with its area, i.e. $\ln t \sim \mathcal{R}^{d-1}$~\cite{Pelissetto:2025pzi,Pelissetto:2026gzg}.

Numerical evidence for this behavior was found prior in temperature-driven first-order transitions of 2D Potts models~\cite{Pelissetto:2016tvy} which led to the following conjectured scaling ansatz for the magnetization
\begin{equation}
    M(J, R) \approx f_M\left(\hat\sigma\right), \quad \text{with } \hat\sigma = (\sigma - \sigma_*)/R^\theta,
    \label{eq:pelissetto_vicari_scaling}
\end{equation}
where $\sigma_*$ represents a common intersection of the magnetization curves when expressed as functions of $\sigma$.
Taking the equilibrium limit $R\to 0$ at fixed $\sigma$ results in the magnetization becoming discontinuous at $\sigma_*$ with corrections to the discontinuity being controlled by the generally temperature-dependent exponent $\theta>0$, which is why this scaling behavior was termed \emph{spinodal-like scaling}~\cite{Pelissetto:2025pzi}.

Although the authors found this scaling ansatz to be in good agreement with results from numerical simulations of the 2D Ising model, as well as simulations of thermal first-order transitions of the two-dimensional $q$-state Potts model with $q=20$ and $q=10$ ~\cite{Pelissetto:2016tvy}, the numerical data for the 3D Ising case was not found to be compatible with the predicted value of $\kappa = 3/2$.
Instead a value of $\kappa = 1$ led to better scaling collapse of the data~\cite{Pelissetto:2025pzi}.
Furthermore, the scaling behavior in three dimensions was observed to be qualitatively different from the two-dimensional case, as the magnetization in this case appeared to scale as a function of $\sigma$ instead of $\hat\sigma$~\cite{Pelissetto:2025pzi}.

\begin{figure}[tb]
    \includegraphics[width=.498\textwidth]{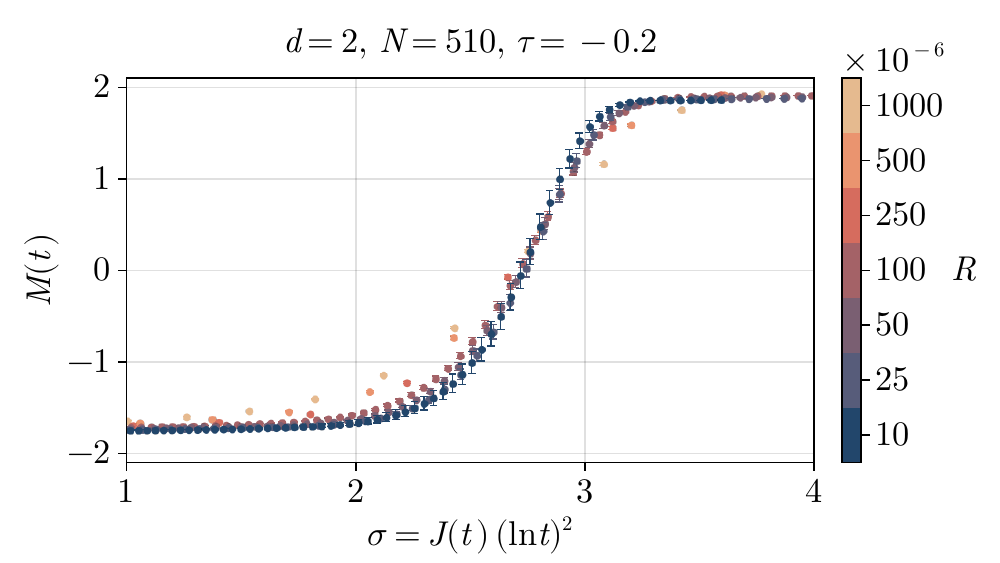}
    \includegraphics[width=.498\textwidth]{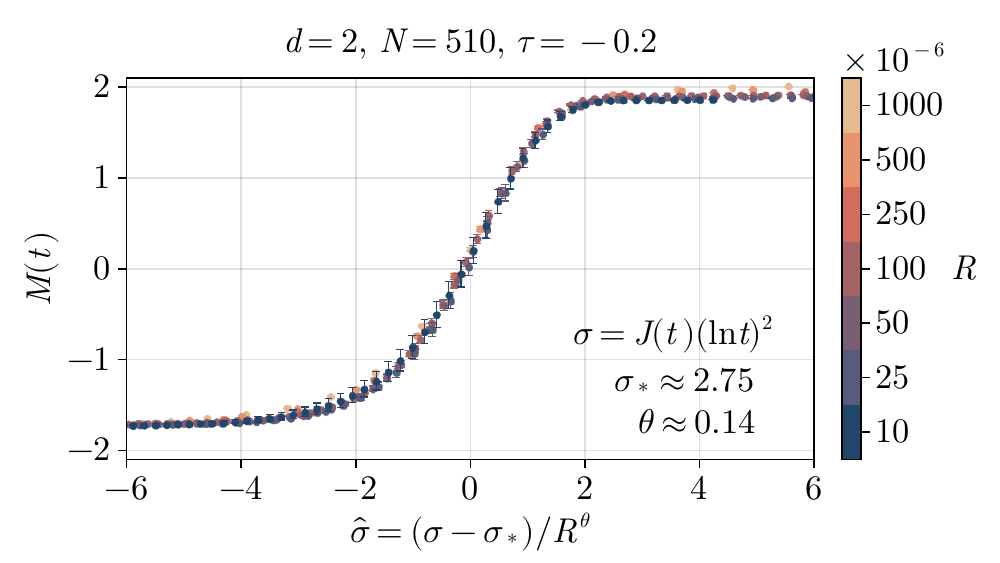}
    \caption{\label{fig:multi_droplet_scaling}
        Order parameter $M$ in $d=2$ spatial dimensions as a function of $\sigma = J(t) (\ln t)^2$ (left) and the scaling variable $\hat\sigma = (\sigma - \sigma_*)/R^\theta$ (right).
    }
\end{figure}

We analyzed our numerical data for evidence of the above scaling behavior in the regime of slow quenches and low temperatures, and were only able to find a limited range of parameters where the scaling behavior in \cref{eq:pelissetto_vicari_scaling} could be observed.
One such example is shown in \cref{fig:multi_droplet_scaling} for the case of $d=2$ spatial dimensions, where we find good scaling collapse of the magnetization curves in terms of the scaling variable $\hat\sigma$ with $\kappa = 2$ for a reduced temperature of $\tau = -0.2$ and quench rates in the range of $ 10^{-5} \lesssim R \lesssim 10^{-2}$.

\begin{figure}[tb]
    \includegraphics[width=.498\textwidth]{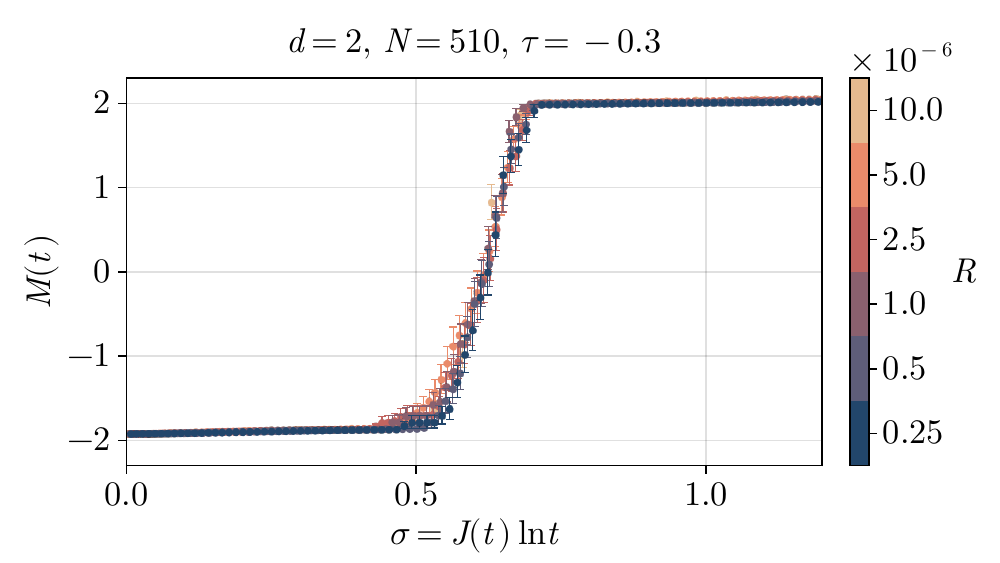}
    \includegraphics[width=.498\textwidth]{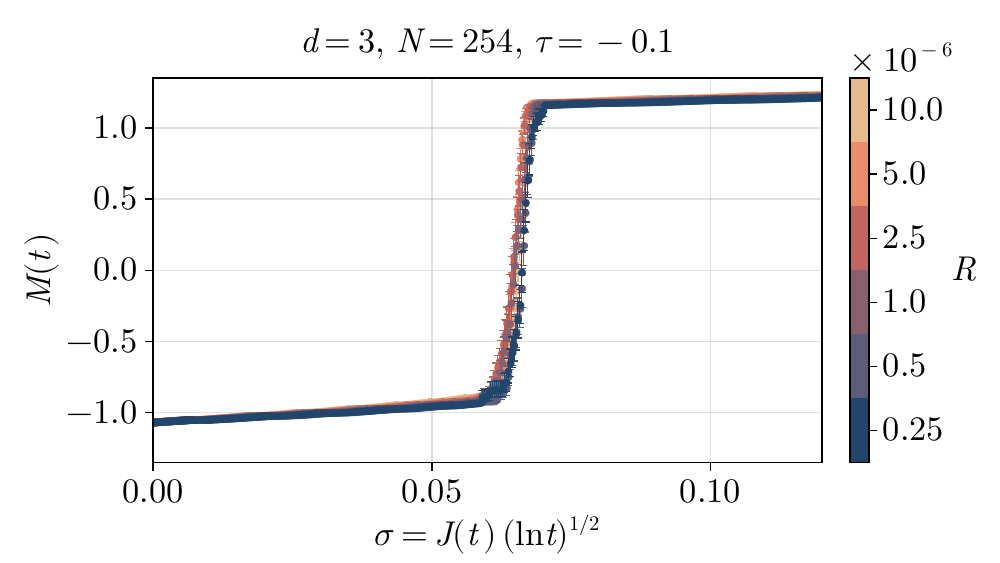}
    \caption{\label{fig:single_droplet_scaling}
        Order parameter $M$ as a function of the scaling variable $\sigma = J(t) (\ln t)^{1/(d-1)}$ in $d=2$ (left) and $d=3$ (right) spatial dimensions.
    }
\end{figure}

For lower temperatures, and slower quenches, where we the transition mechanism should be even more dominated by the nucleation of droplets, we find that the magnetization curves either do not intersect at a single point or do not intersect at all.
Thus, we are not able to find any combination of parameters that lead to a good collapse in terms of the scaling variable $\hat\sigma$ with $\kappa = 2$.
A similar observation was made in~\cite{Zhong:2018}, where it was argued that the proposed logarithmic time factor in the scaling ansatz should only be an approximated form for the behavior of the magnetization.
Though, with $\kappa = 1$ we find in the $d=2$ case apparent scaling in terms of the scaling variable $\sigma$ at a reduced temperature of $\tau = -0.3$ as shown in the left panel of \cref{fig:single_droplet_scaling}.

In the analysis of our $d=3$ data, we were not able to find any temperature where scaling in terms of $\hat\sigma$ or $\sigma$ with $\kappa = 3/2$ or $\kappa=1$ could be observed over a range of quench rates of at least two orders of magnitude.
However, with a value of $\kappa = 1/2$ we found that the $\tau=-0.1$ data showed scaling behavior in terms of $\sigma$ for the slowest quenches we simulated as can be seen in the right panel of \cref{fig:single_droplet_scaling}.
However, just as in two-dimensions, no scaling was observed for significantly lower temperatures.

\begin{figure}[tb]
    \includegraphics[width=.498\textwidth]{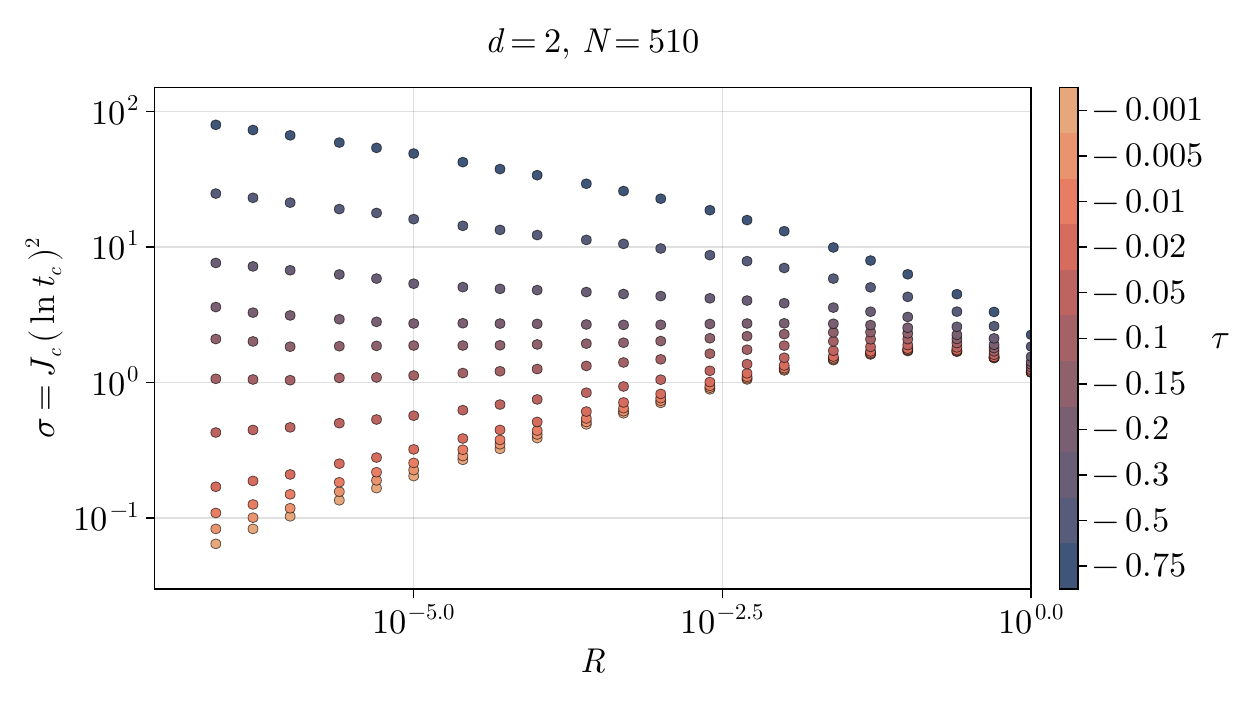}
    \includegraphics[width=.498\textwidth]{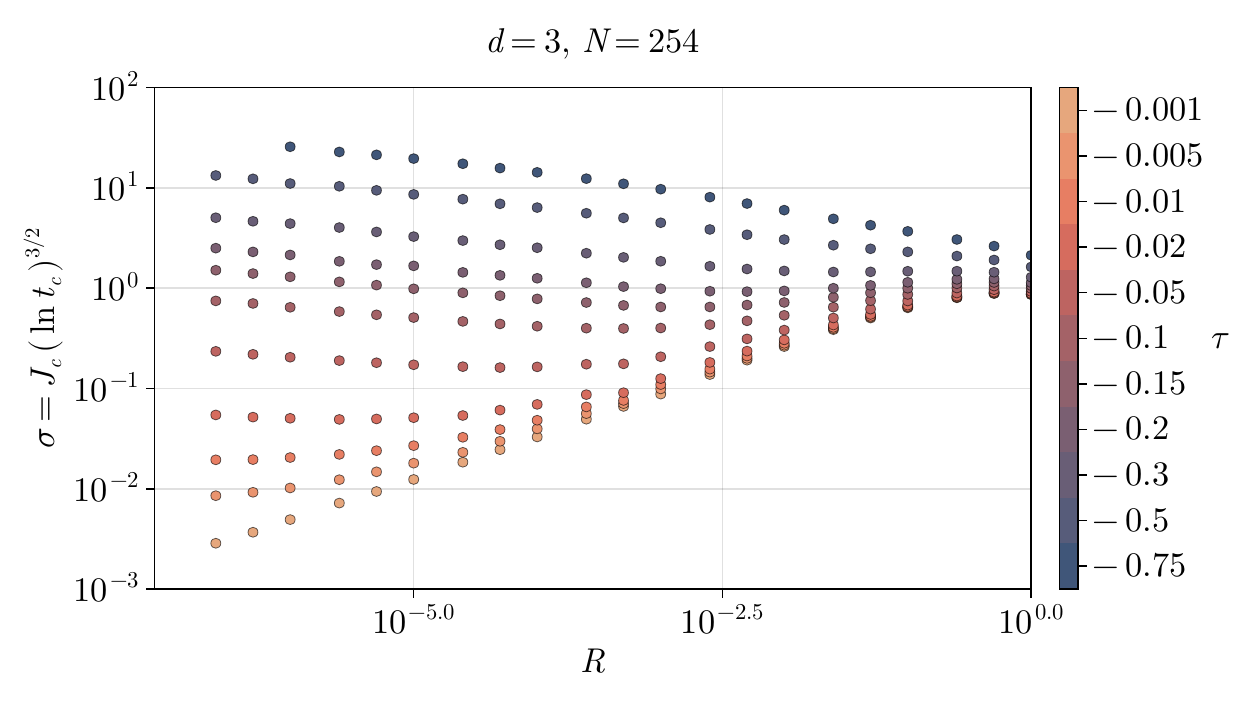}
    \includegraphics[width=.498\textwidth]{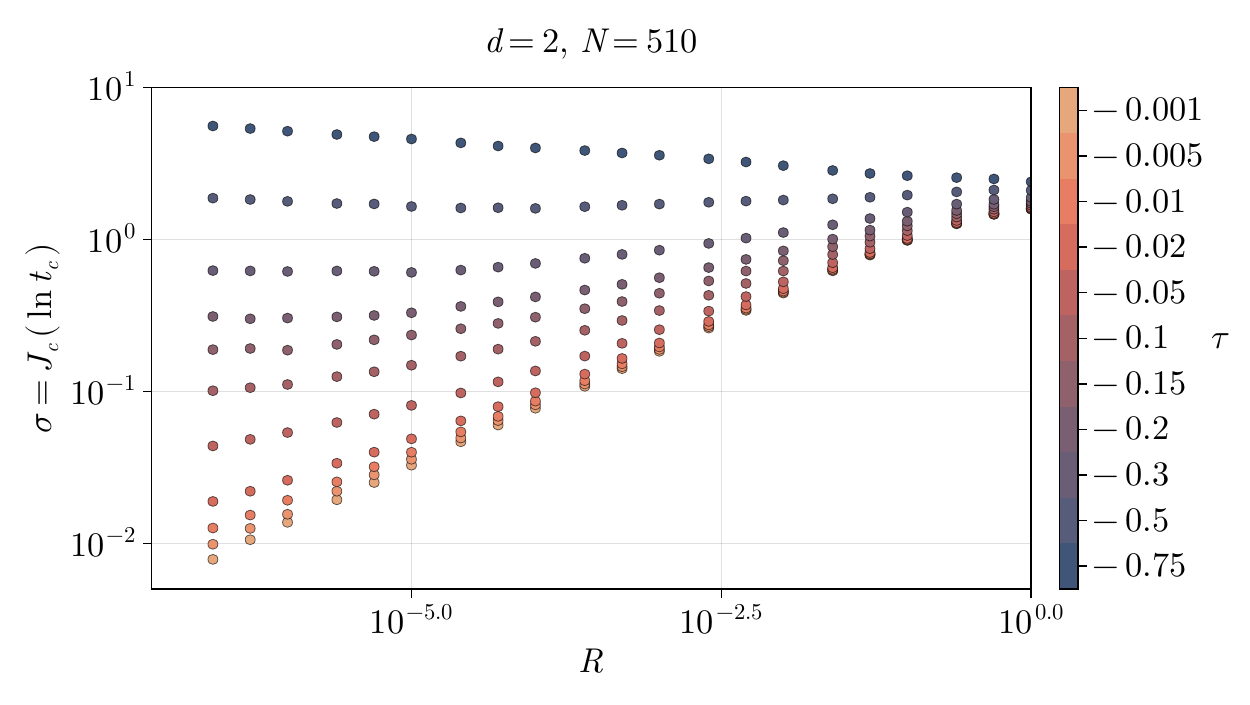}
    \includegraphics[width=.498\textwidth]{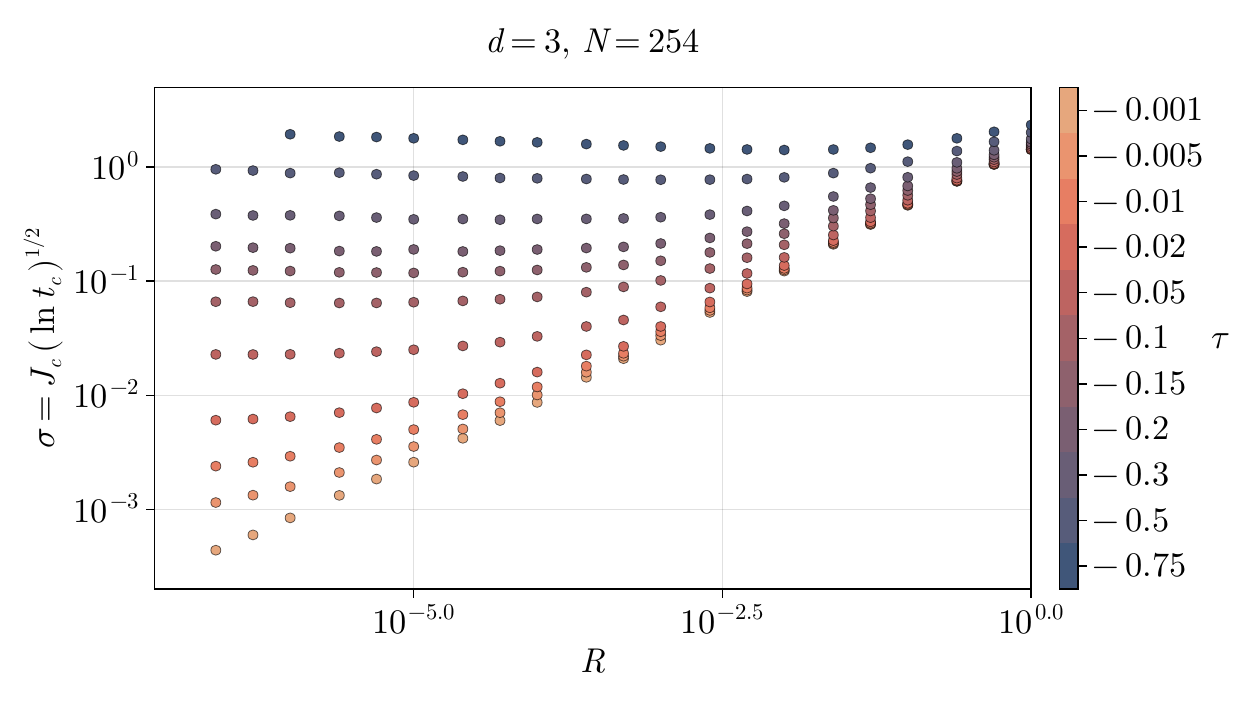}
    \caption{\label{fig:sigma_vs_R} Scaling variable
        $\sigma = J(t) (\ln t)^\kappa$ at the transition point where the magnetization changes sign, plotted as a function of quench rate $R$ for different values of the exponent $\kappa$ in $d=2$ (left) and $d=3$ (right) spatial dimensions.
        The upper panels show the expected scaling variable with $\kappa = d/(d-1)$, while the lower panels show the scaling variable with $\kappa = 1/(d-1)$.
    }
\end{figure}

To be certain that the scaling variable $\sigma$ is applicable close to the transition point, plotting this variable as a function of the quench rate $R$ for a given constant value of $\kappa$ in its definition and a given reduced temperature $\tau$ should result in a plateau over a range of at least a few orders of magnitude in the quench rate $R$.
This is shown in \cref{fig:sigma_vs_R}, where we plot the values of the proposed scaling variable $\sigma$ at the transition point where the magnetization changes sign as a function of the quench rate.
As the value of the external driving field at the transition point is defined to be the coercive field $J_c$, this also defines a coercive time $t_c$ via the relationship $J(t) = R t$.
Accordingly, at the transition point the scaling variable can be obtained to be $\sigma = J_c (\ln t_c)^\kappa$.

In the upper two panels of \cref{fig:sigma_vs_R} we used the expected value of $\kappa = d/(d-1)$ that follows from the general theoretical arguments proposed in~\cite{Pelissetto:2025pzi}, resulting in $\kappa=2$ for the 2D case, shown in upper left panel of \cref{fig:sigma_vs_R} and $\kappa = 3/2$ for the 3D case, shown in the upper right panel.
While in the 2D case with $\kappa=2$ we find a scaling window for a reduced temperature of around $\tau = -0.2$ and quench rates in the range of $ 10^{-5} \lesssim R \lesssim 10^{-2}$, no such scaling window was found in the 3D case with $\kappa =3/2$, as previously mentioned.
In the lower two panels of \cref{fig:sigma_vs_R} we used values of $\kappa$ that resulted in the best plateaus in the regime of asymptotically slow quenches for intermediate temperatures.
These turn out to be $\kappa = 1$ for $d=2$, and $\kappa = 1/2$ for $d=3$ dimensions, although in the $d=2$ case, the plateau is only observed for a small range of temperatures around $\tau = -0.3$, while in three dimensions, the value of $\kappa = 1/2$ seems to be less sensitive to changes in the temperature.

An intriguing observation is that we find scaling in terms of $\sigma$ also in the two-dimensional case with an exponent $\kappa=1$, which was previously only observed in the three-dimensional case~\cite{Pelissetto:2025pzi}.
This may indicate that the underlying transition mechanism is not qualitatively different between two and three spatial dimensions, but rather depends on parameters such as temperature, quench rate and system size.
A natural explanation could be provided by the qualitatively different behavior in multi-droplet and single-droplet nucleation, extensively discussed in the literature~\cite{Tomita:1992,Rikvold:1994za,Sides:1998,Sides:1999}.
In phase transitions dominated by the nucleation and growth of a single droplet, the coercive field is expected to asymptotically behave as $J_c \sim (-\ln R)^{-1/(d-1)}$~\cite{Zhong:2018,Thomas:1993}, which would also be consistent with the exponents of $\kappa=1$ and $\kappa=1/2$ that we find for the slowest quenches in two and three dimensions respectively. At the same time, this scenario does not explain why we fail to observe any clear scaling at the lowest temperatures considered, where the proposed transition mechanism should be dominant. It therefore remains possible that the observed scaling behavior is only apparent.
Further progress would require simulations of significantly larger system sizes in three dimensions, combined with slower quench rates, to unambiguously separate transitions dominated by multi- from single-droplet nucleation and identify possible asymptotic scaling behavior in the respective regimes.

\bibliographystyle{elsarticle-num}
\bibliography{refs}

\end{document}